\documentclass[amsmath,amssymb,prb,groupedaddress,nofootinbib,twocolumn,superscriptaddress, floatfix]{revtex4-2}

\usepackage{times}
\usepackage{changes}

\usepackage{graphicx}

\usepackage{amssymb}
\usepackage{epstopdf}
\usepackage{amsmath}
\usepackage{color}
\usepackage{hyperref}
\usepackage{bbold}
\usepackage{enumitem}
\usepackage[acronym]{glossaries}
\usepackage{physics}
\usepackage[bottom]{footmisc}
\usepackage{txfonts}

\usepackage{slashed}
\usepackage{orcidlink}

\makeatletter

\graphicspath{{Images/}}

\hypersetup{
	pdfpagemode={UseOutlines},
	bookmarksopen,
	pdfstartview={FitH},
	colorlinks,
	linkcolor={black},
	citecolor={blue},
	urlcolor={blue}}

\usepackage{hyperref}
\hypersetup{
	colorlinks=true,
	linkcolor=black,          
	citecolor=blue,           
	filecolor=magenta,        
	urlcolor=cyan
}

\def\120deg{120$^{\circ}$}

\makeatother	

\begin{document}
	\title{Magnetic excitations, phase diagram and order-by-disorder in the extended triangular-lattice Hubbard model}
	
	\author{Josef Willsher \orcidlink{0000-0002-6895-6039}}
 \email{joe.willsher@tum.de}
\affiliation{Department of Physics TQM, Technische Universit{\"a}t M{\"u}nchen, James-Franck-Stra{\ss}e 1, D-85748 Garching, Germany}

	\author{Hui-Ke Jin \orcidlink{0000-0002-0880-6557}}
\affiliation{Department of Physics TQM, Technische Universit{\"a}t M{\"u}nchen, James-Franck-Stra{\ss}e 1, D-85748 Garching, Germany}

	\author{Johannes Knolle \orcidlink{0000-0002-0956-2419}}
\affiliation{Department of Physics TQM, Technische Universit{\"a}t M{\"u}nchen, James-Franck-Stra{\ss}e 1, D-85748 Garching, Germany}
\affiliation{Munich Center for Quantum Science and Technology (MCQST), 80799 Munich, Germany}
\affiliation{\small Blackett Laboratory, Imperial College London, London SW7 2AZ, United Kingdom}

	\begin{abstract}
The dynamical structure factor is an important observable of quantum magnets but due to numerical and theoretical limitations, it remains a challenge to make predictions for Hubbard-like models beyond one dimension. In this work, we study the magnetic excitations of the triangular lattice Hubbard model including next-nearest neighbor hopping. Starting from the expected {\120deg} and stripe magnetic orders we compute the magnon spectra within a self-consistent random phase approximation. In the stripe phase, we generically find accidental zero modes related to a classical degeneracy known from the corresponding $J_1$--$J_2$ Heisenberg model. We extend the order-by-disorder mechanism to Hubbard systems and show how quantum fluctuations stabilize the stripe order. In addition, the frustration-induced condensation of magnon modes allows us to map out the entire phase diagram which is in remarkable agreement with recent numerical works. We discuss connections to experiments on triangular lattice compounds and the relation of our results to the proposed chiral spin liquid phase.
\end{abstract}
	
\maketitle

\section{Introduction} 
The Hubbard model has played a central role in condensed matter research as one of the simplest effective models describing interacting electrons~\cite{hubbard1963electron,arovas2022hubbard}. It displays a whole variety of fascinating many-body phenomena such as high-temperature superconductivity, Mott metal-insulator transitions, and different forms of quantum magnetism~\cite{auerbach2012interacting}. 
In one dimension, the ground state and excitations of the Hubbard model have been understood in great detail~\cite{Lieb1968,Giamarchi2003,essler2005one}. However, in the absence of exact methods, in two dimension and beyond the ground state phase diagram remains a challenge even for state-of-the-art  numerical methods~\cite{hirsch1985two,white1989numerical,leblanc2015solutions}, with access to the excitaion spectrum proving even more difficult.
In addition, on geometrically frustrated lattices the complexity is even higher ~\cite{Vannimenus1977,Toulouse1987,BookDiep,BookLacroix} since competing magnetic interactions provide a potential for realizing quantum spin liquid (QSL) phases~\cite{Anderson73,Anderson87,Lee08,Balents10,Savary2016,QSLRMP,Knolle2019,Broholm2020}. In particular, the successful synthesis of possible QSL candidates~\cite{li2020spin} on the triangular lattice, including organic compounds $\kappa$(ET)$_2$Cu$_2$(CN)$_3$ and Me$_3$EtSb[Pd(dmit)$_2$]$_2$~\cite{Kanoda2011,Powell2011}, or rare-earth compounds YbMgGaO$_4$~\cite{Li2015,Shen2016} and NaYbO$_2$~\cite{ding2019gapless} has triggered a lot of activity. More recently, the rapid progress in the field of twisted moir\'e materials~\cite{Cao2018,Cao2018_2,Yankowitz2019,Wu2018,Wu2019} provides a versatile platform for the study of Hubbard physics on various lattices and with different band structures~\cite{Wang2020,Kennes2021,arxiv:2209.05506}.
In particular, the triangular-lattice Hubbard model can be realized in twisted bilayer boron nitride~\cite{Ni2019,Xian2019}, twisted WSe$_2$/WS$_2$ moir\'e superlattices~\cite{Tang2020,Regan2020}, and twisted double-bilayer WSe$_{2}$~\cite{An2020}.

Much progress has been made in the study of the Hubbard model on the triangular lattice by using combinations of theory and numerics. 
In the large-$U$ limit and with only nearest neighbor (NN) hopping $t$, the Hubbard model at half filling is reduced to the antiferromagnetic Heisenberg model for which the celebrated resonating valence bond (RVB) state was proposed as a trial wave function by Anderson in 1973~\cite{Anderson73} --- nowadays considered to be the first QSL state. 
However, the following numerical studies provided clear evidence for a long-range {\120deg} antiferromagnetic (AFM) order in the strong coupling limit~\cite{Huse1988,Bernu1994,Capriotti1999,White2007,LiQian2022}. Away from strong coupling, a spin stiffness analysis based on a mean-field approximation indicates a loss of {\120deg} magnetic order at $U\sim{}6t$~\cite{PhysRevB.71.214406}. 
An intermediate QSL phase (the so-called weak-Mott insulator) might additionally be stabilized for moderate Hubbard interaction $U\sim{}8t$, as indicated by the variational cluster approximation (VCA)~\cite{Sahebsara2008}, path-integral renormalization group analysis~\cite{Yoshioka2009}, and series expansions~\cite{Yang2010}. Moreover, recent density matrix renormalization group (DMRG) studies give strong evidence for a gapped chiral QSL in this regime \cite{Shirakawa2017,Szasz2020,Wietek2021,PhysRevB.103.235132,zhu2022doped,chen2022quantum,PhysRevLett.127.087201}.

The Hubbard model with additional next-nearest-neighbor (NNN) hoppings $t^\prime$ is further frustrated and studies have mostly been restricted to the large-$U$ model, i.e., the $J_1$--$J_2$ AFM Heisenberg model on the triangular lattice. In the classical limit where the spins are treated as $O(3)$ vectors, it is predicted that the coplanar {\120deg} order is destabilized and a highly degenerate four-sublattice state appears between $1/8\leq{}J_2/J_1\leq{}1$~\cite{PhysRevB.42.4800}; for larger values of $J_2/J_1>1$ spiral states dominate. The intermediate regime has been one of the paradigmatic examples of the \textit{order-by-disorder} mechanism~\cite{villain1980order,henley1989ordering} where quantum fluctuations treated within spin-wave theory lift the classical degeneracy such that a collinear stripe ordered phase is stabilized~\cite{PhysRevB.42.4800,Chubukov1992}. 
Near the classical critical point at $J_2/J_1=1/8$ an exciting possibility is the emergence of a magnetically disordered phase, such as the $U(1)$ Dirac QSL state which has been proposed within the variational Monte Carlo (VMC) approach~\cite{Kaneko2014,Iqbal2016}. DMRG works have suggested gapped \cite{Zhu2015,Hu2015} and more recently gapless \cite{Gong2017, Hu2019,drescher2022dynamical} QSL ground states.

Only few works have studied the full Hubbard model with both the NN and NNN hopping at finite $U$ including charge and spin fluctuations. A ground-state phase diagram has been obtained in the $t'$-$U$ plane by using a VCA method with a 12-site cluster~\cite{Misumi2017} and VMC~\cite{Tocchio2020}, which together point to a rich phase diagram including insulating spin disordered phases, similar to other extensions of the Hubbard model including e.g. anisotropic hopping~\cite{doi:10.1143/JPSJ.71.2109,PhysRevB.77.214505,PhysRevB.80.064419,PhysRevB.87.035143,PhysRevB.103.235132,arxiv.2211.09234}.
However, theoretical studies of the dynamical spin excitations as well as an understanding of the metal-insulator and magnetic phase transitions are missing, despite being of particular importance in light of the recent above-mentioned candidate materials.

Despite recent breakthroughs in numerical techniques which allow ever more detailed studies of Hubbard-like models, strong limitations on system size and frequency resolution remain for the calculation of dynamical properties in two dimensions. Specifically, DMRG studies of two-dimensional magnets are limited in their prediction of dynamical correlation functions due to entanglement growth in time.
Recent effort has begun to shine light on the dynamical properties of the strong coupling Heisenberg limits~\cite{drescher2022dynamical,sherman2022spectral}, but it remains challenging to probe Hubbard physics in the full $t/U$ regime \cite{PhysRevB.106.094417}. In this work we provide calculations of dynamical properties of the triangular lattice $t$--$t'$--$U$ Hubbard model. Working within a self-consistent mean field theory we are able to calculate the dynamical spin-spin correlation function within the random phase approximation (RPA). Our method, which includes  harmonic spin and charge fluctuations around the magnetically ordered states, has been pioneered for the square lattice Hubbard model~\cite{PhysRevB.39.11663,PhysRevB.46.11884,PhysRevB.41.614} and successfully applied to describe inelastic neutron scattering experiments for cuprates~\cite{peres2002spin,peres2003spin} and iron-based superconductor~\cite{knolle2010theory,brydon2009spin,kaneshita2010spin,knolle2011multiorbital} parent compounds.

\begin{figure*}
\centering
\includegraphics[width=\textwidth]{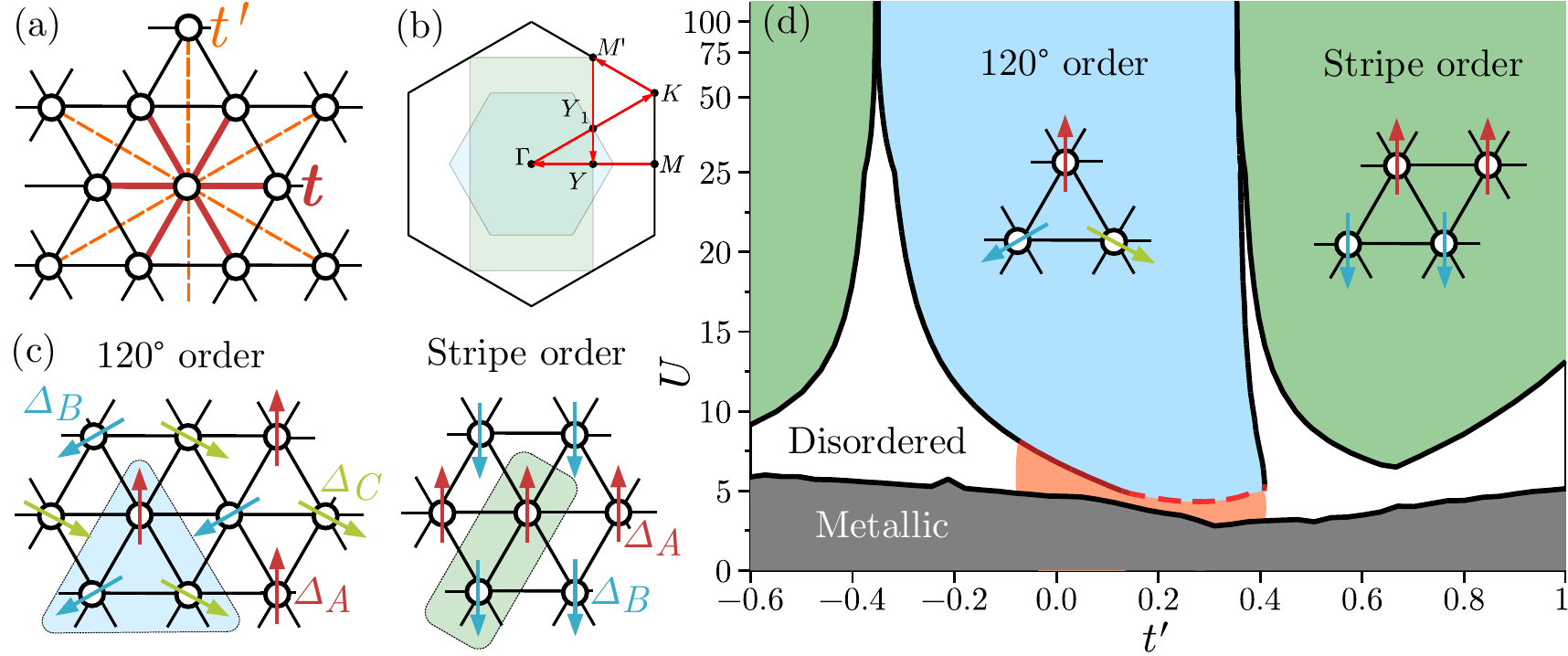}
\caption{\textbf{Model and Phases}: (a) Triangular lattice with nearest neighbor $t$ (solid) and next-nearest neighbor $t'$ hoppings (dashed line) highlighted. (b) Brillouin zone showing the highlighted magnetic Brillouin zones for each ordered phase. The path drawn is used to plot spectral functions. 
(c) {\120deg} (left) and stripe (right) ordered states. Unit cells highlighted in grey boxes and magnetic moments $\Delta_\alpha$ highlighted for each order. 
(d) Phase diagram of the triangular lattice Hubbard model, calculated using mean field theory plus fluctuations with on-site repulsion $U$. The phases highlighted are: central {\120deg} order (blue); peripheral stripe order (green); at low $U$ a metallic phase (gray); intermediate insulating spin-disordered phases (white); and a potential noncoplanar insulating state (pink).
The solid boundaries of ordered phases highlight where short-wavelength fluctuations lead to a transition into a spin-disordered phase.
The dashed boundary of the {\120deg} phase is a transition driven by vanishing transverse spin-wave velocity; red boundary coloring shows the suppression of this quantity extends down to $t'\approx-0.1$.}
\label{diagram}
\end{figure*}

We find a large region with self-consistent solutions for both the {\120deg}- and stripe-ordered phases. 
Indeed both competing solutions are allowed for much of the area above the metallic phase marked in Fig.~\ref{diagram}(d) (including all $t'$ for $U\gtrsim10$).
Instead of comparing the respective ground state energies to produce a pure mean-field phase diagram, we use the calculation of spin excitations to examine the stability of the mean-field magnetization in the presence of fluctuations. 
Most of the phase boundaries of the {\120deg} and stripe phases can be mapped out by observing the condensation of \emph{accidental soft modes} when the magnon spectrum comes down to zero-energy at a wavevector incommensurate with the existing magnetic order.
We additionally find a novel phase boundary at small positive $t'$ and $U$ where magnetic order is instead destabilized directly by a vanishing of the spin wave velocity. From the components of the structure factor we observe that the corresponding fluctuations are out-of-plane indicating an instability to a phase with non-collinear spin correlations. Of course, our method is only able to describe excitations within a long-range magnetically ordered phase but we argue that the unusual out-of-plane instability is an indication of the nearby chiral spin liquid phase.

In addition to the phase diagram, we provide a detailed understanding of the magnetic excitations in the stripe phase. Similar to the corresponding $J_1$--$J_2$ model, for a stripe state with wave vector $M$ we find within our Hubbard model approach accidental soft modes at $M'$ as a direct signature of the underlying magnetic degeneracy for sizable $t'$. We then show that the inclusion of quantum fluctuations via an approximate treatment of self-energy effects gaps out the accidental mode at $M'$ stabilizing the collinear stripe phase. Thus, we extend the order-by-disorder mechanism to the triangular lattice Hubbard model.

Our manuscript is structured as follows: In Section~\ref{sec:mft} we present the model and its mean-field solution. In Section~\ref{sec:spin} we explain the calculation of the magnetic phases' spin fluctuations in the RPA scheme. In Section~\ref{sec:obd} we discuss in detail the stability of the stripe ordered phase beyond harmonic fluctuations; and in Section~\ref{sec:transitions} discuss how phase transitions of the model are diagnosed and used to map out the phase diagram. We conclude with a discussion of our results and open questions.

\section{Mean field theory}\label{sec:mft}
The Hubbard model Hamiltonian is given by
\begin{equation}\label{eq:H}
    H = -\sum_{i,\delta}t_\delta \, a^\dagger_{i\sigma}a_{i+\delta\sigma} + U\sum_i a^\dagger_{i \uparrow}a_{i\uparrow} \,a^\dagger_{i\downarrow}a_{i\downarrow},
\end{equation}
where $\delta$ is a nearest or next-nearest neighbor displacement vector on the triangular lattice, as shown in Fig.~1(a), and $a^\dagger_{i\sigma}$ is a creation operator of a fermion with spin $\sigma$ at site $i$. The (next) nearest neighbor hopping is henceforth referred to as $t$ ($t'$). The on-site repulsion is $U$ and the chemical potential is fixed to half filling.

First, we review the standard self-consistent mean-field theory for the two different magnetic orderings on the triangular lattice: {\120deg} order at the $K$ point in the Brillouin zone (BZ); and stripe order at the $M$ point.
Fig.~1(b) shows the first BZ of a triangular lattice with the ordering vectors $K$ and $M$ highlighted, alongside other high-symmetry points and a path between them which is used for our analysis of dynamical spectral functions across the BZ. In the presence of symmetry breaking and magnetic order formation, the magnetic unit cell is enlarged; it contains $N_{\mathrm{UC}}=3$ sites for the {\120deg} and $N_{\mathrm{UC}}=2$ sites for stripe order with the magnetic ordering pattern highlighted in Fig.~1(c).
The respective magnetic BZ (MBZ) is shrunk covering 1/3 and 1/2 of the BZ area [Fig.~1(b)]. One may consider equivalent stripe orderings at the other midpoints of the BZ edge (labeled $M'$), which are related by a sublattice-dependent spin rotation and are part of the underlying classical manifold of magnetic states~\cite{Chubukov1992}.

With these magnetic states we perform a Hartree--Fock (HF) decoupling of the Hamiltonian Eq.~\eqref{eq:H} which is implemented by introducing \emph{sublattice fermions} $a_\alpha$, where the sublattice index $\alpha=A,B,C$ for the {\120deg} Ansatz and $\alpha=A,B$ for the stripe Ansatz. 
The mean-field Hamiltonian elements are hence labeled by sublattice index $\alpha$ in addition to spin indices $\sigma = \uparrow,\downarrow$.

The ordering is encoded in this formalism with the sublattice vectors $\vec{\Delta}_\alpha$ (components $(\Delta_\alpha)_{\mu}$, for $\mu = x,y,z$).
In our calculations, we choose the following basis for mean-field vectors as highlighted in real space in Fig.~1(c): the {\120deg} state has $\vec{\Delta}_A =  (0,\Delta,0)$ with $\vec{\Delta}_{B,C}$ rotated clockwise by $120$ and $240$ degrees; the stripe state has $\vec{\Delta}_{A,B} = \pm(0,0,\Delta)$, where we have used the fact that the mean-field collinear state is independent of the choice of ordering direction to fix it out of the plane. 
Written in this compact form, the decoupled Hamiltonian is
\begin{equation}
    H_{\mathrm{MF}} = \sum_{i,\alpha} a_{i,\alpha\sigma}^\dagger [-(\Delta_\alpha)_\mu (\sigma^\mu)_{\sigma\sigma'} ] a_{i, \alpha \sigma'} + \sum_{i,\delta} t_\delta \, a^\dagger_{i,\sigma}a_{i+\delta,\sigma} ,
\end{equation}
where the repeated spin indices are implicitly summed over.
In addition to this, the sublattice magnetization vector is defined self-consistently 
\begin{equation}\label{eq:sublmag}
    \Delta_{\alpha,\mu} = U \left<(S_\alpha)_\mu\right>_{\mathrm{GS}} = \frac{U}{2} \left< a_{\alpha\sigma}^\dagger (\sigma_\mu)_{\sigma\sigma'} a_{\alpha\sigma} \right>_{\mathrm{GS}}
\end{equation}
where $\vec{S}_\alpha$ is the sublattice spin vector and expectation value $\left<~\cdot~\right>_{\mathrm{GS}}$ is with respect to the ground state.
The sublattice magnetization $m = 2\Delta/U$ is calculated self consistently for each state.

We next diagonalize the system by Fourier-transformation; introducing the sublattice spinor $\Psi_{\mathbf{k}} = ([a_{\mathbf{k}}]_{\uparrow,A}, [a_\mathbf{k}]_{\downarrow,A},\dots)^T$ with 6 (or 4) compound indices $m=(\alpha,\sigma)$ for {\120deg} (stripe) sublattice fermions, we can rewrite the Hamiltonian as
$H_{\mathrm{MF}} = \sum_{\mathbf{k}} \Psi^\dagger_{\mathbf{k}} H(\mathbf{k}) \Psi_{\mathbf{k}}$. Explicitly the mean-field Hamiltonian in the stripe phase is therefore
\begin{equation}
    H_{\mathrm{stripe}}(\mathbf{k}) =
\begin{pmatrix}
\zeta_{\mathbf{k}}\mathbb{I}-\sigma\cdot \Delta_A & \delta_{\mathbf{k}}\mathbb{I} \\
\delta_{\mathbf{k}}^*\mathbb{I} & \zeta_{\mathbf{k}}\mathbb{I}-\sigma\cdot \Delta_B
\end{pmatrix},
\end{equation}
with $\mathbb{I}$ the identity matrix and $AA,BB$ and $AB,BA$ hoppings written as $\zeta_{\mathbf{k}}$ and $\delta_{\mathbf{k}}$, respectively. These are calculated by summing over all $AA$ or $AB$ hopping vectors $\delta$ for both NN and NNN bonds, defined in the following way:
\begin{equation}\label{eq:sumofhoppings}
    \delta_{\mathbf{k}} = \sum_{\delta=A\to B} t_{\delta} e^{i \mathbf{k}\cdot \delta},
    \quad
    \zeta_{\mathbf{k}} = \sum_{\delta=A\to A} t_{\delta} e^{i \mathbf{k}\cdot \delta}.
\end{equation}
Explicit expressions for these matrix elements are given in the Appendix alongside the equivalent formulation for the {\120deg} mean-field Hamiltonian.

Diagonalizing the HF Hamiltonian produces eigenstates $\ket{\mathbf{k},m}$ of the Hamiltonian with spectrum $\varepsilon_m(\mathbf{k})$, where $m=1\dots 2N_{\mathrm{UC}}$. Mean field expectation values can be computed in this basis, including the ordering vector via Eq.~\eqref{eq:sublmag}; in a system of linear dimension $L$, its expectation value is given
\begin{align}\label{eq:evdelta}
    \Delta_{\alpha,\mu} &= \frac{U}{2} \left< \Psi^\dagger \sigma^\mu_\alpha\Psi \right>_{\mathrm{GS}} \nonumber\\
    &= \frac{U}{2N} \sum_{\mathbf{k},m} n_f[\varepsilon_{m}(\mathbf{k})-\mu] \bra{\mathbf{k},m} \sigma_\alpha^\mu \ket{\mathbf{k},m}
\end{align}
where $N = L^2 N_{\mathrm{UC}}$ is the number of sites and $n_f[\varepsilon_{m}(\mathbf{k})-\mu]$ is the Fermi--Dirac distribution at zero temperature. The sublattice-$\alpha$ Pauli matrix $\sigma_\alpha^\mu$ is defined as acting on a $\Psi^{\sigma'}_\beta$ object through the Pauli matrix $(\sigma^\mu)_{\sigma\sigma'}$ on its spin index and on its sublattice index diagonally via $\delta^{\alpha\beta}$.
At half-filling, the chemical potential $\mu(U,\Delta)$ is always fixed such that the fermion occupation
\begin{equation}
    n = \left< \Psi^\dagger\Psi \right>_{\mathrm{GS}} = \frac{1}{N} \sum_{\mathbf{k},m} n_f[\varepsilon_{m}(\mathbf{k})-\mu]
\end{equation}
is equal to $1$ per site.

Numerically, the magnetic order $\Delta$ is computed self consistently as follows: for a given choice $\Delta$ the mean-field Hamiltonian is found, then the chemical potential $\mu$ is fixed by minimizing $n-1$ numerically, and finally the expectation of magnetic ordering [found via Eq.~\eqref{eq:evdelta}] is computed. The difference $\left< \Psi^\dagger \sigma^\mu_\alpha\Psi \right>_{\mathrm{GS}} - 2\Delta/U$ is then minimized numerically to self-consistently compute $\Delta$.

\begin{figure*}
\begin{minipage}{.49\linewidth}
\centering
\includegraphics[width=\textwidth]{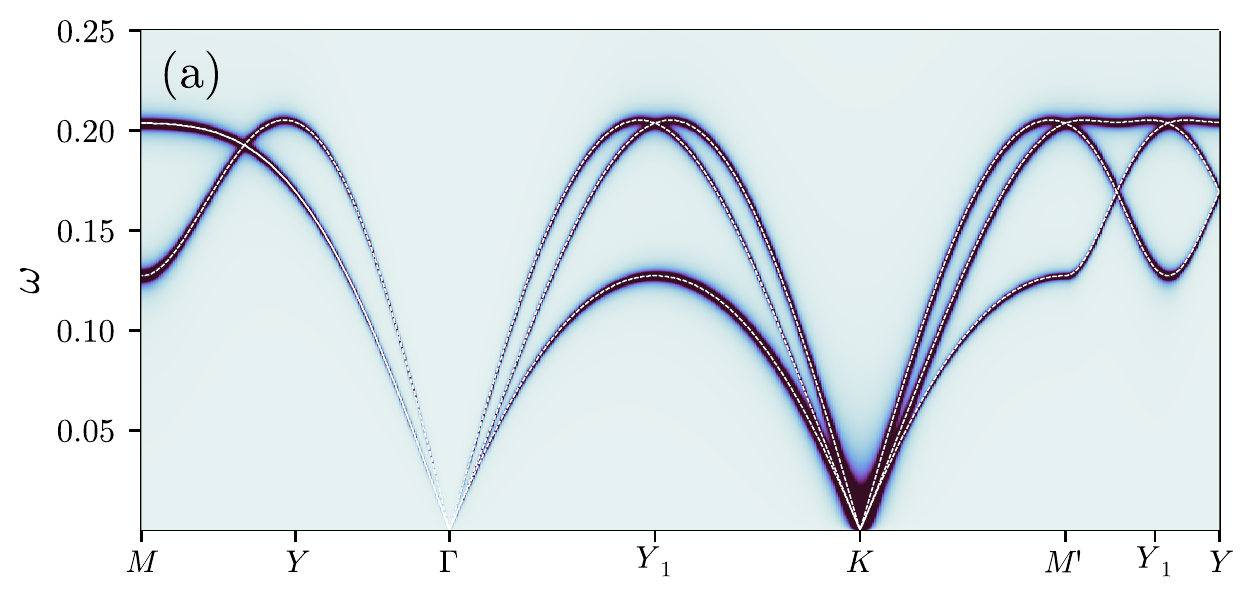}
\end{minipage}
\begin{minipage}{.49\linewidth}
\centering
\includegraphics[width=\textwidth]{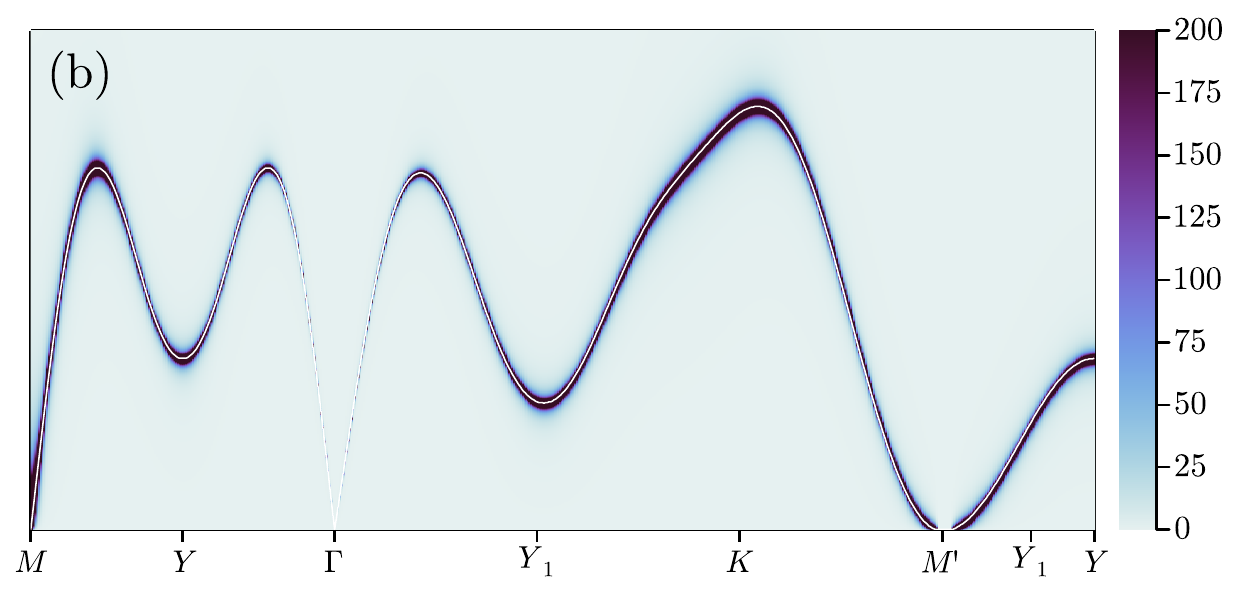}
\end{minipage}
\caption{\textbf{Dynamical structure factor} $S(\mathbf{q},\omega)$ of (a) the {\120deg} and (b) stripe magnetic ordered phases, taken along the path in the Brillouin zone highlighted in the text. White lines represent the zeros of $[1 - U \chi^{(0)}(\mathbf{q},\omega)]$, solid lines highlight doubly degenerate modes. Systems $U=30$, linear size $L=720$ ($N= N_{\mathrm{UC}} L^2$ sites), with (a) $t'=0$ and (b) $t'=0.6$.}
\label{spectra}
\end{figure*}

\section{Spin fluctuations}\label{sec:spin}
The spin susceptibility is one of the prime observables for understanding the physics of Hubbard systems: This quantity gives access to the \emph{dynamical structure factor}, an experimentally observable quantity which is routinely probed in inelastic neutron  scattering experiments~\cite{lovesey1984theory}. 
As Goldstone modes in an AFM system, the spin waves extend down to low energy with linear dispersion, and understanding the velocity of these modes at low energy allows the prediction of quantum corrections to the sublattice magnetization. Access to the dynamical spin susceptibility will hereby provide information about phase boundaries beyond the basic mean-field approximation.
The high-energy spectrum of spin excitations may in particular show strong deviations from Heisenberg-like spin-only physics, due to the effect of charge fluctuations especially in the intermediate-$U$ regime.

The dynamical spin susceptibility $\chi(\mathbf{q},\omega)$ can be expressed in terms of the momentum-space spin-spin correlation function~\cite{moriya2012spin},
\begin{equation} 
    [\chi(\mathbf{q},\omega)]^{\mu\nu}_{\alpha\beta}
    = \frac{i}{2 N} \int \dd{t} \, e^{i\omega t} \left< T S_{\mathbf{q},\alpha}^\mu(t) S_{-\mathbf{q},\beta}^\nu(0)\right>_{\mathrm{GS}},
\end{equation}
defined with sublattice indices $\alpha,\beta$ and spin indices $\mu,\nu$.
Explicitly, the spin operators can be written as $S_{\mathbf{q},\alpha}^\mu = \sum_{\mathbf{k}} \Psi_{\mathbf{k}+\mathbf{q}}^\dagger [\sigma^\mu_\alpha] \Psi_{\mathbf{k}}$.

In order to calculate this susceptibility, we focus on the magnetic interaction channel and apply the RPA formalism \cite{PhysRevB.39.11663,PhysRevB.46.11884,knolle2011multiorbital}. The susceptibility tensor is defined through the Dyson equation
\begin{equation}
     [\chi]^{\mu\nu}_{\alpha\beta} = [\chi]^{\mu\rho}_{\alpha\gamma} U^{\rho\sigma}_{\gamma\lambda} [\chi^{(0)}]^{\sigma\nu}_{\lambda\beta} + [\chi^{(0)}]^{\mu\nu}_{\alpha\beta},
\end{equation}
where the interactions are contained in the tensor $U$. This is diagrammatically expressed as
\begin{equation}
\begin{gathered}
\includegraphics[scale=0.66]{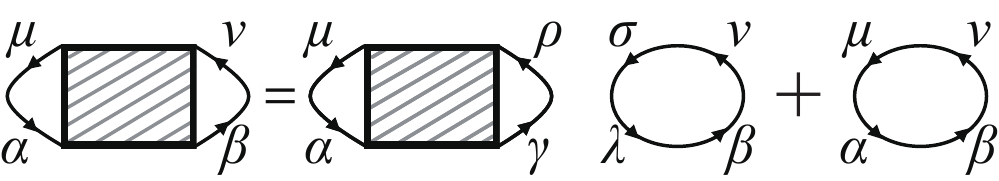}
\end{gathered}
\end{equation}
in terms of the non-interacting correlation function $\chi^{(0)}(\mathbf{q},\omega)$.
The interaction vertex is taken to be diagonal in both indices
\begin{equation}
U^{\rho\sigma}_{\gamma\lambda} = U \mathbb{I} = 
\begin{gathered}
\includegraphics[scale=0.66]{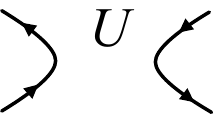}\,.
\end{gathered}
\end{equation}
The solution to the equation can be conveniently written by summing up all bubble diagrams \cite{PhysRevB.41.614,PhysRevB.43.3617}; in matrix form this is
\begin{equation}\label{chifull}
     \chi(\mathbf{q},\omega) = \left[\mathbb{I} - U \chi^{(0)}(\mathbf{q},\omega)\right]^{-1} \chi^{(0)}(\mathbf{q},\omega).
\end{equation}
Our task therefore becomes to evaluate this bare correlation function using the mean-field state; it can be expressed as
\begin{align}
    [\chi(\mathbf{q},\omega)^{(0)}]^{\mu\nu}_{\alpha\beta}
=
\frac{1}{N}
\sum_{\mathbf{k}}&\sum_{m,l}^{2N_{\mathrm{UC}}} \{n_f[\varepsilon_{m}(\mathbf{k})]-n_f[\varepsilon_{l}(\mathbf{k}+\mathbf{q})]\}
\nonumber \\ \times &
\frac{[\mathcal{V}_{ml}(\mathbf{k})]^\nu_\alpha
[\mathcal{V}^*_{lm}(\mathbf{k}+\mathbf{q})]^\mu_\beta}{\omega+\varepsilon_{m}(\mathbf{k})-
\varepsilon_{l}(\mathbf{k}+\mathbf{q})-i\delta},
\end{align}
where $[\mathcal{V}_{ml}(\mathbf{q})]^\mu_\alpha$ is the following matrix element of the spin-operator
\begin{equation}
    [\mathcal{V}_{ml}(\mathbf{q})]^\mu_\alpha = \left<\mathbf{q},m\right|\sigma^\mu_\alpha\left|\mathbf{q},l\right>.
\end{equation}
The effect of increasing $N$ is to improve the resolution in the MBZ by reducing the finite-size level spacing.
The divergence of the susceptibility is regulated by a width $i\delta$, which we taken proportional to $t^2/(NU)$ so as to scale with the level spacing in a finite system and to smoothly take the thermodynamic and strong coupling limits.
    
The bare correlation function has poles corresponding to particle-hole excitations, which are gapped for half-filling in all magnetic phases. The collective modes are contained in the zeros of $[1 - U \chi^{(0)}(\mathbf{q},\omega)]$ corresponding to the spin wave excitations in a fully self-consistent treatment. 
In order to reproduce the dynamical spectral function, we calculate the full matrix $\chi(\mathbf{q},\omega)$ and then evaluate the following sum over spins and sublattices
\begin{equation}
    S(\mathbf{q},\omega) = \operatorname{Im} \sum_{\alpha,\beta}^{N_{\mathrm{UC}}}\, e^{-i\mathbf{q}\cdot (r_\alpha - r_\beta)} \sum_{\mu,\nu=x,y,z}  [\chi(\mathbf{q},\omega)]^{\mu\nu}_{\alpha\beta}.
\end{equation}
Moreover, the eigenvalues of $1 - U \chi^{(0)}(\mathbf{q},\omega)$ are evaluated for each wavevector $\mathbf{q}$, the zeros of which define the spin-wave dispersion $\omega_{\mathbf{k}}$.

Fig.~\ref{spectra}(a) shows a typical structure factor $S(\mathbf{q},\omega)$ evaluated in the {\120deg} ordered phase away from the phase boundaries. The ordering wavevector is at the $K$ points in the BZ, and spectral weight vanishes at the $\Gamma$ point. This is in agreement with previous RPA results~\cite{PhysRevB.71.214406} (in which only the $t'=0$ case was studied), as well as recent numerical studies of the {\120deg}-ordered phase of the Heisenberg limit \cite{drescher2022dynamical,sherman2022spectral}.
This coplanar but non-collinear state has three spin excitation modes, corresponding to eigenvectors of the susceptibility matrix; at low energies these can be characterized as either purely in-plane ($S^x,S^y$) or out-of-plane ($S^z$).
Note that magnon excitations encoded in the susceptibility are still transverse to the local ordering vector on each sublattice. 
We distinguish their nature by calculating $\operatorname{Im}\chi^{zz}$ for out-of-plane modes, and $\operatorname{Im}\chi^{+-}$ for in-plane modes (with $\left<S^+ S^-\right>$ containing fluctuations in $S^x$ and $S^y$). 
These Goldstone modes have a linear dispersion at low $\omega$, where the single in-plane mode has a higher velocity than the two degenerate transverse modes $c_{\parallel} > c_{\perp}$.

Looking to the high-energy behavior, the in- and out-of-plane modes generally mix character but remain degenerate across the $M\to Y\to \Gamma$ line and at the other high symmetry points $Y_1, M'$ where the highest-energy bands meet and where the fluctuations are predominantly in-plane.

Turning to the stripe ordered phase, Fig.~\ref{spectra}(b) shows a typical structure factor $S(\mathbf{q},\omega)$ away from the phase boundaries. The ordering wavevector here is at the $M$ point in the BZ with linear dispersion; a similar linear mode also appears at the $\Gamma$ point but with vanishing spectral weight. There is only one excitation mode in this collinear state which is twofold degenerate everywhere in the BZ.
The ordering vector is chosen out of plane, and so $\operatorname{Im}\chi^{+-}$ contains all information about the magnon dispersion.

A further advantage of the self-consistent RPA approach is that it allows the calculation of dynamical response functions away from half filling. The method naturally includes charge fluctuations from particle-hole excitations which cannot be captured within a spin-only treatment. We have checked that both the {\120deg} and stripe phases are stable against hole doping; the two-particle continuum is visible as weak spectral weight, and there is an increase in the spin-wave velocity for low doping.
We also confirmed a strong particle-hole asymmetry, as was seen for the $t'=0$ model in Ref.~\cite{PhysRevB.71.214406}, but since a full study of the doping dependent phase diagram is beyond the scope of this work we concentrate on the case of half-filling for the remainder.

\section{Order-by-Disorder in the Stripe Phase}\label{sec:obd}
Turning back to the stripe phase spectrum in Fig.~\ref{spectra}(b), one can observe a peculiar mode at the point $M'$ which goes quadratically to zero despite the magnetic ordering wavevector being at the inequivalent $M$ point.
In the large-$U$ limit, the $t,t'$ Hubbard model becomes the corresponding frustrated Heisenberg $J_1$--$J_2$ model, whose stripe order has an exact \emph{accidental zero mode}
according to linear spin wave theory (LSWT)~\cite{Chubukov1992,PhysRevB.42.4800}. 
Taken at face value, the zero-energy spin excitations at this point with zero stiffness would produce a divergent quantum correction to the sublattice magnetization, destroying the stripe state, thus, at the harmonic level no long-range order is selected. 
However, this is an artifact of the quadratic-fluctuation approximation; in the Heisenberg model it is known that quantum fluctuations generically gap the accidental zero mode at $M'$ to higher energies --- the gap becomes of the order $J\sim t^2/U$ --- leading to a stable phase while the Goldstone modes at the $M$ point remain gapless~\cite{Chubukov1992}. The general scenario that quantum (or thermal or quenched disorder) fluctuations stabilize a long-range ordered state is known as \emph{order-by-disorder}.

\begin{figure}[h]
\centering
\includegraphics[width=\columnwidth]{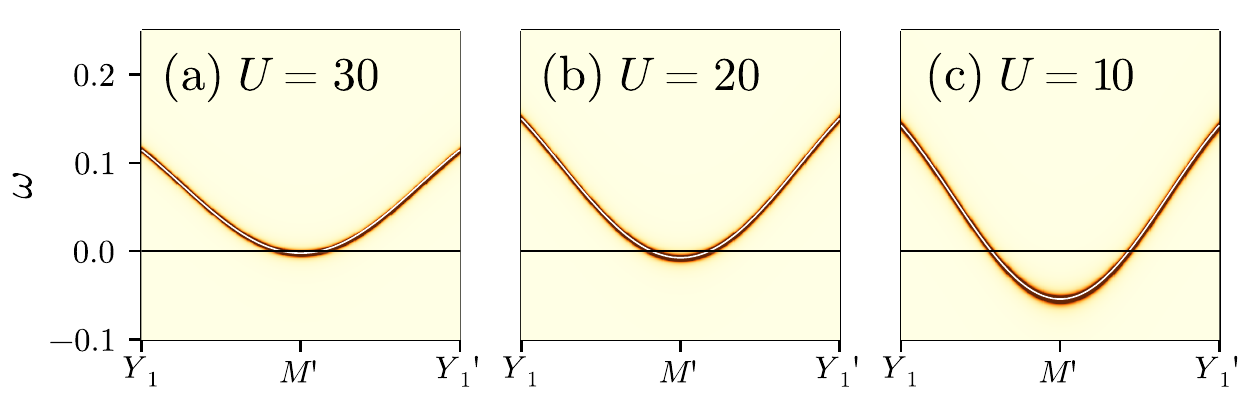}
\includegraphics[width=\columnwidth]{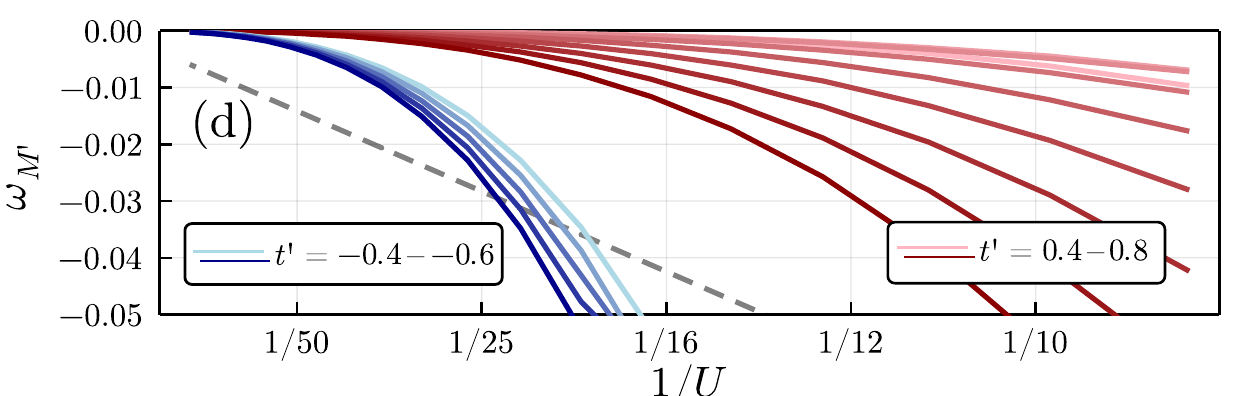}
\caption{\textbf{Stripe phase instability in the RPA scheme} of the accidental zero mode around the $M'$ point. Subfigures (a)--(c) plot the transverse susceptibility $\operatorname{Im}\chi^{+-}(\mathbf{q},\omega)$ for $U=30$, $20$, and $10$. The white line is the doubly degenerate mode $\omega_{\mathbf{q}}$ calculated from the zeros of $[1-U\chi^{(0)}]$. Subfigures (a)--(c) show data calculated with parameter $t'=0.8$ and linear size $L=720$ ($N=2 L^2$ sites). Subfigure (d) shows the negative energy of the mode minimum $\omega_{M'}$ for varying values of $t'$ across the phase diagram, indicating a failure of the RPA. The red/blue lines (color online) are positive/negative values of $t'$, with darker lines having a higher $|t'|$. The dashed line shows the (negative) magnitude to the Heisenberg-limit’s fluctuation-induced gap $J \sim t^2/U$.}
\label{mppoint}
\end{figure}

Turning back to the Hubbard model's stripe phase in the RPA approximation, we find that reducing $U$ actually tends to lower the minimum of the mode $\omega_{M'}$ below zero energy, clearly indicating a failure of the random phase approximation.
Fig.~\ref{mppoint}(a)--(c) plots the transverse susceptibility $\operatorname{Im}\chi^{+-}$ for varying $U$, showing that the minimum is significantly below zero energy at $U\approx 15$ with positive $t'$ (and happens sooner for negative $t'$).
In Fig.~\ref{mppoint}(d) this is put quantitatively, showing the $U^{-3}$ dependence of this minimum energy. This becomes of equal magnitude to the Heisenberg-limit's fluctuation-induced gap for negative $t'$ around $U\approx 20$. Here, one may expect that spin-only fluctuations may not restore stripe-order even beyond the RPA scheme.

The presence of the accidental zero mode at $M'$ is an artifact of the harmonic fluctuation approximation scheme (here the RPA), and will not hold up in the presence of quantum fluctuations. Indeed, the square lattice antiferromagnetic and ferromagnetic phases were studied in Refs.~\cite{PhysRevB.43.3617, edwards1973electron}, where quantum fluctuations were calculated systematically for the Hubbard model by extending it to contain $\mathcal{N}$ orbitals.
By evaluating the fermion self-energy to order $1/\mathcal{N}$, the effect on the susceptibility was shown to be a shifted pole with renormalized weight. 
In the following, we provide an approximate treatment of quantum fluctuations around the $M'$ point on the triangular lattice stripe phase and show that the accidental mode remains gapped down to low $U$.
We use the fact that the effect of quantum fluctuations on $\chi$ can be calculated by replacing $\chi^{(0)}$ in Eq.~\eqref{chifull} with $\phi = \chi^{(0)}+\delta\phi$, where the additional term originates from diagrams which appear alongside the free loop $\chi^{(0)}$ in the Dyson equation~\cite{PhysRevB.43.3617}.

\begin{figure}[h]
\centering
\includegraphics[width=\columnwidth]{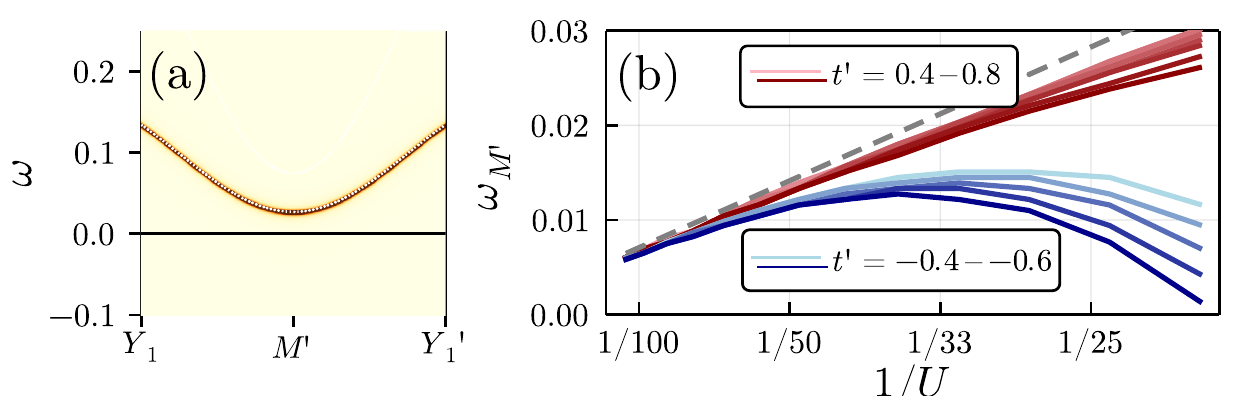}
\caption{\textbf{Stripe phase stability including quantum fluctuations} of the accidental zero mode around the $M'$ point. Subfigure (a) plots the transverse susceptibility $\operatorname{Im}\chi^{+-}(\mathbf{q},\omega)$ for $U=20$. The white dashed line is the mode $\omega_{\mathbf{k}}$ calculated from the zeros of $[1-U\phi]$. Figure shows data calculated with parameter $t'=0.8$ and linear size $L=720$ ($N=2 L^2$ sites). Subfigure (b) shows the energy of the mode minimum $\omega_{M'}$ for varying values of $t'$ including the phenomenological self-energy correction. The red/blue lines (color online) are positive/negative values of $t'$, with darker lines having a higher $|t'|$. Scaling at high $U$ is linear, with the dashed line $\propto t^2/U$.}
\label{mppointselfen}
\end{figure}

We can calculate an approximate form of $\phi$, generally given by
\begin{equation}
    \delta\phi(\mathbf{k}, \omega) = \frac{1}{\mathcal{N}}\frac{t^2}{U^3}
\begin{pmatrix}
K_{\mathbf{k}} +J_{\mathbf{k}}  \, U\omega/2t^2 & K_{\mathbf{k}}\gamma_{\mathbf{k}}  \\
K_{\mathbf{k}} \gamma_{\mathbf{k}} & K_{\mathbf{k}} - J_{\mathbf{k}}\, U\omega/2t^2\label{deltaphi}
\end{pmatrix}
\end{equation}
where $K_{\mathbf{k}}$, $J_{\mathbf{k}}$ are factors coming from loop diagrams~\cite{PhysRevB.43.3617}. 
These are momentum dependent and of the order $1/\mathcal{N}$ in the controlled expansion parameter.
The term $\gamma_{\mathbf{k}}$ is the off-diagonal element of $\chi^{(0)}$, which is zero when evaluated at the $M'$ point in the RPA.
The loop-functions $K_{\mathbf{k}}, J_{\mathbf{k}}$ are not able to be constrained in this way and therefore are expected to be order one. 
We may now form an approximation of Eq.~\ref{deltaphi}, valid for momenta close to $M'$, by approximating these functions as constants: $\gamma_{\mathbf{k}}=0$ and $K_{\mathbf{k}} = J_{\mathbf{k}} \sim 1$.
In Fig.~\ref{mppointselfen}(a) we show that indeed the spectrum is corrected to be gapped at the $M'$ point  (for concreteness we used $U=20$, $t=0.8$). Thus, we find that similar to the Heisenberg limit, fluctuations also stabilize the triangular lattice Hubbard model's stripe phase.

In this corrected picture and for large on-site repulsion $U$, we find that the gap magnitude scales as $\omega_{M'} \propto t^2/U$ in agreement with the Heisenberg limit~\cite{Chubukov1992} (the exact constant relating the gap to $J$ in this limit depends on the values $J_{M'}$, $K_{M'}$). 
For positive $t'$ this stability extends down to below the phase transition to a spin-disordered phase, suggesting the order-by-disorder effect stabilizes the whole stripe phase here.

For negative $t'$, the $\omega_{M'}$ gap appears to vanish again at around $U\approx 20$, where the fluctuation-correction $\propto t^2/U$ becomes comparable to the $\omega_{M'}$ of the quadratic RPA-theory, see Fig.~\ref{mppointselfen} [and the dashed line in Fig.~\ref{mppoint}(d)].
We suspect that in this regime either the phase becomes disordered, or that diagrams involving charge fluctuations become significant and may further stabilize the phase.
One could further improve our approximate treatment by including higher-order self-energy corrections and by calculating these self-consistently~\cite{PhysRevB.77.094430} which comes at a considerable numerical cost. This is beyond the scope of our work but  we argue that our results provide good evidence that for much of the coupling regime $U\gtrsim10$ supporting stripe ordering, the magnon spectrum at the $M'$ point is gapped. In the following, we will therefore ignore the artificial zero mode at the $M'$ point within the harmonic approximation and instead focus on the softening of collective modes at other points in the BZ for mapping out the phase diagram.

\section{Phase Transitions \& Discussion}\label{sec:transitions}
We can now map out the phase diagram of the next-nearest neighbor triangular lattice Hubbard model, going beyond mean-field theory by requiring stability against fluctuations. 
Our main result is shown in Fig.~\ref{diagram}(d) which is in remarkable in agreement with numerical works based on the variational cluster approximation~\cite{Misumi2017} and more recently on the variational Monte Carlo method~\cite{Tocchio2020}. We find the presence of {\120deg} order at low $|t'|$, stripe order for increased $|t'|$, and instabilities towards spin-disordered phases in the intermediate-$U$ regime. At sufficiently low $U$ there is a metallic phase characterized by no self-consistent magnetic order and a gapless spectrum.

Despite only considering quadratic fluctuations in the RPA scheme, we observe two types of phase transition out of the ordered phases: Firstly, the magnon modes come down to zero energy with a linear dispersion around some incommensurate critical wavevector; Secondly, the spin-wave velocity around the ordering wavevector vanishes. We will discuss the physical implications of these two transitions in the following. These two types of transitions can be easily characterized, in contrast to the descending accidental zero mode at $M'$ discussed in the previous section, which does not show any critical behaviour (remaining quadratic as it descends down in energy at the RPA level). 

\begin{figure}[h!]
\centering

\includegraphics[width=\columnwidth]{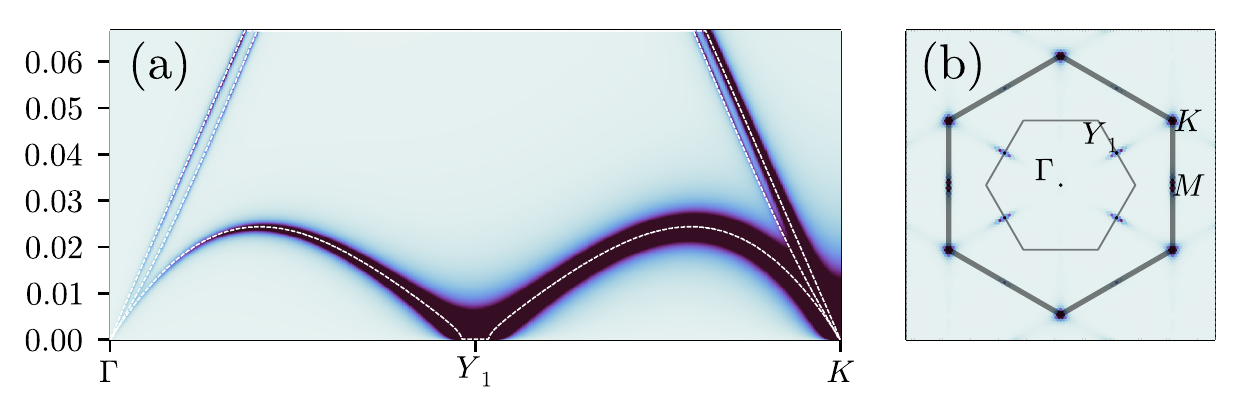}
\includegraphics[width=\columnwidth]{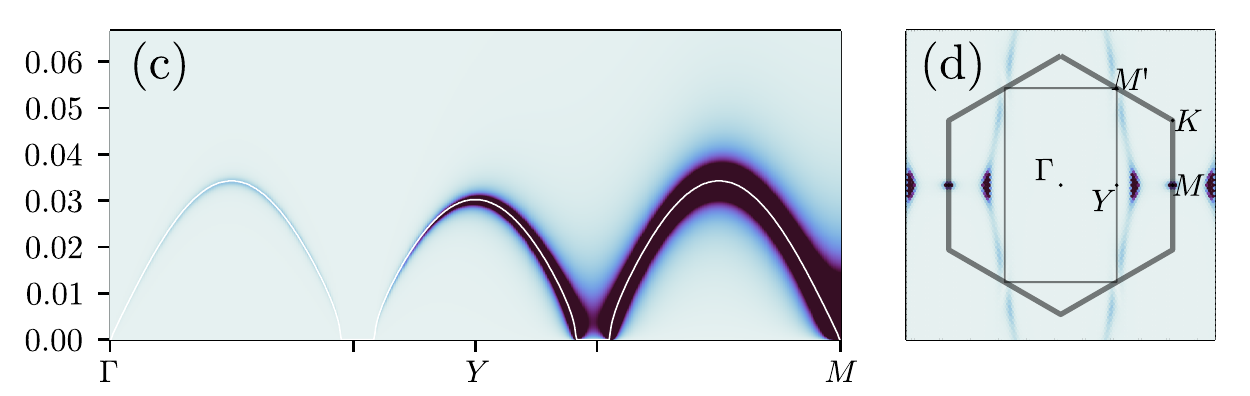}

\caption{\textbf{Critical modes closing} for $U=30$ with (a,b) $t'=0.358$ in the {\120deg} phase and (c,d) $t'=0.365$ in the stripe ordered phase. The figures (a,c) show the dynamical structure factor $S(\mathbf{q},\omega)$ along the paths from $\Gamma$ to the respective ordering wavevector. White lines represent the zeros of $[1 - U \chi^{(0)}(\mathbf{q},\omega)]$, solid lines highlight doubly degenerate modes. The static structure factor $S(\mathbf{q})$ is plotted in (b,d). Linear size $L=720$ ($N= N_{\mathrm{UC}} L^2$ sites) for spectral functions and $L=72$ for static structure factor.}
\label{closingfigs}
\end{figure}

The first type of transition occurs when the ordered phase is destabilized by a magnon branch reaching zero energy at a wavevector incommensurate with the ordering vector.
In this regime, despite a self-consistent mean-field solution, the Ansatz is not stable against fluctuations.
We find that the {\120deg} phase has a transition of this type whereby a transverse spin fluctuation mode closes at the $M$ point (and equivalently the $Y_1$ point in the MBZ) see Fig.~\ref{closingfigs}(a,b). 
This is shown as solid lines in Fig.~\ref{diagram}(d).
This was previously observed along the $t'=0$ line for varying $U$ \cite{PhysRevB.71.214406}, but we find the behaviour persists along the whole extent of the phase boundary.

Numerically the boundary is ascertained by examining two features, which happen simultaneously: the diverging matrix elements $\chi(\mathbf{q}_{\mathrm{closing}},0)$ at zero energy which become larger than the same elements at the ordering wavevector, as well as the lowest zero eigenvalue of $[1-U \chi(\mathbf{q}_{\mathrm{closing}},\omega)^{(0)}]$ (the mode energy) reaching zero $\omega_{\mathrm{closing}} = 0$. 
Passing through the phase boundary, these modes become linear and then square-root like, with no pole in the imaginary part of the susceptibility matrix at the critical closing point.

The RPA scheme can easily be extended to the $U\to\infty$ limit where it predicts the direct transition between {\120deg} and stripe ordered phases at $t'_c \approx 0.35$. 
At strong coupling the system also appears symmetric in $t'$, and from the {\120deg} phase the closing wavevector is at the $M$ point.
In connection to the classical results for the $J_1$--$J_2$ Heisenberg model, the exchanges to leading order in perturbation theory are $J_1 =4t^2/U$ and $J_2 = 4t'^2/U$.
Our results predict in this limit $J_{2,c} \approx 0.123$, in good agreement with the classical $J_{2,c} = 1/8$ \cite{PhysRevB.42.4800,Chubukov1992}. 

From the stripe ordered phase however, the closing wavevector in the RPA scheme is dependent on the position along the phase boundary; at high-$U$ the closing wavevector is at the corner of the three-site {\120deg} unit cell MBZ, see Fig.~\ref{closingfigs}(c,d). At reduced $U$ towards the bottom of the lobe of stripe order in the phase diagram Fig.~\ref{diagram}(d), the closing vector moves towards $Y$ and then to $K$. This is in the region where the RPA scheme results in a negative-energy accidental mode, and we expect that the position of this closing vector could be moved in a self-consistent calculation which considers higher-order fluctuations.

\begin{figure}[h]
\centering
\includegraphics[width=\columnwidth]{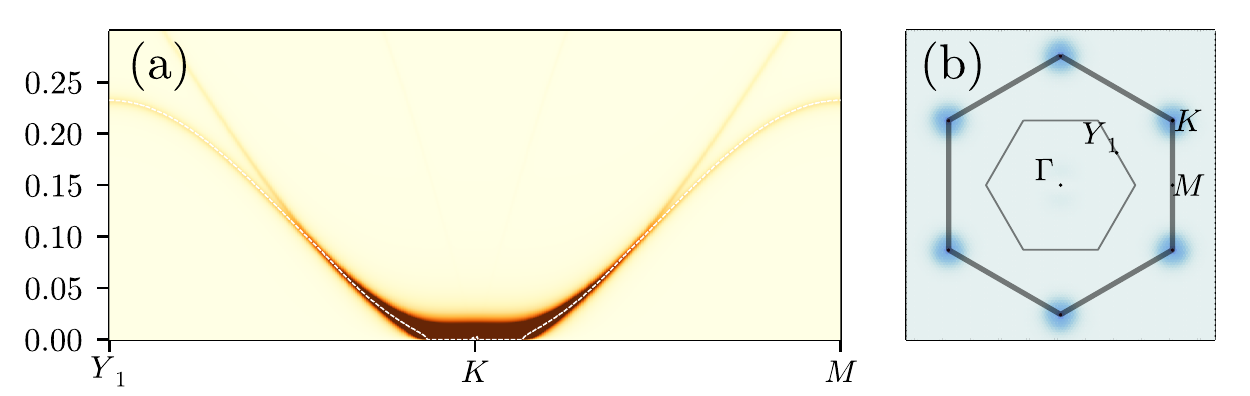}
\caption{\textbf{Vanishing of the spin wave velocity} along the red transition line of the phase diagram. 
(a) The transverse susceptibility $\operatorname{Im}\chi^{zz}(\mathbf{q},\omega)$. The dashed line represents the lowest zero of $[1 - U \chi^{(0)}(\mathbf{q},\omega)]$. (b) The static structure factor $S(\mathbf{q})$.
Calculations for $U=4.6$, $t'=0.289$; 
linear size $L=720$ ($N= 3 L^2$ sites) for dynamical susceptibility and $L=72$ for static structure factor.}
\label{softmode}
\end{figure}

The second type of transition is triggered by long-wavelength spin excitations. When the spin-wave velocity of excitations about the ordering wavevector goes to zero, the quantum correction to the sublattice magnetization are expected to diverge~\cite{Chubukov1992}.
This is shown as dashed lines in Fig.~\ref{diagram}(d), where we find that the transverse velocity going to zero preempts the closing of the mode at $M$ for finite $t'\approx 0.15$--$0.4$ in the {\120deg} ordered phase. Therefore, the order is directly destroyed by long-wavelength transverse fluctuations in this regime, marked as a red line in Fig.~\ref{diagram}(d). Fig.~\ref{softmode} shows the transverse susceptibility components, and highlights the massless out-of-plane Goldstone excitations at this transition.
We also observe for even lower $t'$ down to approximately $-0.1$ that the spin-wave velocity at the transition point is strongly suppressed, as indicated by the red-coloured boundary region in Fig.~\ref{diagram}(d). This could indicate the presence of a transverse-fluctuation-driven transition over a wider range of the colored phase boundary at low $U$ \cite{PhysRevB.71.214406}.

It is tempting to relate this transition to the recent numerical works observing a chiral QSL for $t'=0$ and small $U$~\cite{Shirakawa2017,Szasz2020,Wietek2021}. Indeed, the strong out-of plane fluctuations would be naturally related to an instability towards a non-coplanar magnetically ordered state, which has been proposed as the parent state of the chiral QSL~\cite{PhysRevB.96.115115,saadatmand2017detection,gong2017global} and the soft structure factor at this point is also observed numerically in the CSL phase~\cite{PhysRevLett.127.087201}.
Of course, we cannot make definite claims about the nature of the disordered phases because our RPA scheme assumes the presence of a magnetic state. Nevertheless, our results certainly highlight the different nature of the order-disorder transition from the {\120deg} phase along the bottom and sides of the phase boundary. 
At higher $U$ there is a clear tendency to melting because of short-wavelength fluctuations. At lower $U$ there is a regime where out-of-plane fluctuations destabilize the order directly (for $0.15\lesssim t' \lesssim 0.4$), as well as a larger region encompassing the $t'=0$ line where the transverse spin stiffness may also be sufficiently suppressed to destroy the sublattice magnetization.

\section{Conclusion}
We have studied the spin fluctuations of the $t$--$t'$--$U$ Hubbard model on the triangular lattice and mapped out the phase diagram at half filling. Employing a self-consistent mean-field plus RPA approximation allowed us to calculate the spin structure factor in the thermodynamic limit. Therefore, the method is ideally suited to provide predictions for inelastic neutron scattering experiments complementary to the usual spin wave calculations applicable only in the insulating large interaction limit.

Despite its limitations, we have shown that the method provides a comprehensive picture of the spin fluctuations in the ordered stripe and {\120deg} phases. Thereby, it allows the delineation of phase boundaries according to the stability of magnon modes. We find that the phase diagram shown in Fig.~\ref{diagram}(d) is in remarkable agreement with recent numerical works. In general, the self consistent RPA theory is well suited to the study of itinerant magnetic systems, including away from half-filling. In the future it would be worthwhile to explore the entire doping dependence of the triangular lattice Hubbard model. 

We have provided a first study of spin excitations within the stripe magnetically ordered phase. We showed how the inclusion of quantum fluctuations removes an accidental zero mode from the underlying magnetic degeneracy, which provides an extension of the order-by-disorder mechanism to the finite-$U$ regime~\cite{PhysRevB.42.4800,Chubukov1992}.
This was done by modifying the bare susceptibility to account for a fluctuation-renormalized self-energy~\cite{PhysRevB.43.3617, edwards1973electron}. In the future it would be interesting to include higher-order corrections and perform a fully self-consistent calculation. While technically and numerically challenging, this would provide a more complete picture of the effect of charge-fluctuations and the stability of magnetic phases for hole and electron doping.

The phase boundaries in the large $U$ limit are in agreement with a classical transition between ordered phases at $J_2/J_1 = 1/8$ \cite{PhysRevB.42.4800,Chubukov1992}. In the finite $U$ regime we also find spin disordered regimes which have been at the centre of recent numerical works both for the corresponding Heisenberg limit~\cite{White2007,Hu2015,Iqbal2016} and for the Hubbard model~\cite{Shirakawa2017,Szasz2020,Wietek2021}. Given the difficulty of accessing with purely numerical methods the dynamical behavior of two-dimensional frustrated models with QSL ground states, a complementary approach is provided by the self-consistent RPA method which has been extended to parton-mean-field descriptions of Heisenberg models with fractionalized excitations~\cite{ho2001nature,zhang2020resonating}. An exciting direction for future research is to explore similar schemes for QSL phases of Hubbard models in order to understand the nature of spin excitations in the presence of spin fractionalization and charge fluctuations.

\section*{Acknowledgements}
We would like to thank Philipp Gegenwart, Markus Drescher and Frank Pollmann for useful discussions. J.K. acknowledges support from the Imperial-TUM flagship partnership. The research is part of the Munich Quantum Valley, which is supported by the Bavarian state government with funds from the Hightech Agenda Bayern Plus. H.-K.J. is funded by the European Research Council (ERC) under the European Unions Horizon 2020 research and innovation program (Grant agreement No.~771537).

\textbf{Data and materials availability} – Code and data are available on Zenodo \cite{willsher_josef_2022_7290111}.

\bibliographystyle{apsrev4-2}
\bibliography{main}

\begin{thebibliography}{96}%
\makeatletter
\providecommand \@ifxundefined [1]{%
 \@ifx{#1\undefined}
}%
\providecommand \@ifnum [1]{%
 \ifnum #1\expandafter \@firstoftwo
 \else \expandafter \@secondoftwo
 \fi
}%
\providecommand \@ifx [1]{%
 \ifx #1\expandafter \@firstoftwo
 \else \expandafter \@secondoftwo
 \fi
}%
\providecommand \natexlab [1]{#1}%
\providecommand \enquote  [1]{``#1''}%
\providecommand \bibnamefont  [1]{#1}%
\providecommand \bibfnamefont [1]{#1}%
\providecommand \citenamefont [1]{#1}%
\providecommand \href@noop [0]{\@secondoftwo}%
\providecommand \href [0]{\begingroup \@sanitize@url \@href}%
\providecommand \@href[1]{\@@startlink{#1}\@@href}%
\providecommand \@@href[1]{\endgroup#1\@@endlink}%
\providecommand \@sanitize@url [0]{\catcode `\\12\catcode `\$12\catcode
  `\&12\catcode `\#12\catcode `\^12\catcode `\_12\catcode `\%12\relax}%
\providecommand \@@startlink[1]{}%
\providecommand \@@endlink[0]{}%
\providecommand \url  [0]{\begingroup\@sanitize@url \@url }%
\providecommand \@url [1]{\endgroup\@href {#1}{\urlprefix }}%
\providecommand \urlprefix  [0]{URL }%
\providecommand \Eprint [0]{\href }%
\providecommand \doibase [0]{https://doi.org/}%
\providecommand \selectlanguage [0]{\@gobble}%
\providecommand \bibinfo  [0]{\@secondoftwo}%
\providecommand \bibfield  [0]{\@secondoftwo}%
\providecommand \translation [1]{[#1]}%
\providecommand \BibitemOpen [0]{}%
\providecommand \bibitemStop [0]{}%
\providecommand \bibitemNoStop [0]{.\EOS\space}%
\providecommand \EOS [0]{\spacefactor3000\relax}%
\providecommand \BibitemShut  [1]{\csname bibitem#1\endcsname}%
\let\auto@bib@innerbib\@empty
\bibitem [{\citenamefont {Hubbard}(1963)}]{hubbard1963electron}%
  \BibitemOpen
  \bibfield  {author} {\bibinfo {author} {\bibfnamefont {J.}~\bibnamefont
  {Hubbard}},\ }\href@noop {} {\bibfield  {journal} {\bibinfo  {journal}
  {Proceedings of the Royal Society of London. Series A. Mathematical and
  Physical Sciences}\ }\textbf {\bibinfo {volume} {276}},\ \bibinfo {pages}
  {238} (\bibinfo {year} {1963})}\BibitemShut {NoStop}%
\bibitem [{\citenamefont {Arovas}\ \emph {et~al.}(2022)\citenamefont {Arovas},
  \citenamefont {Berg}, \citenamefont {Kivelson},\ and\ \citenamefont
  {Raghu}}]{arovas2022hubbard}%
  \BibitemOpen
  \bibfield  {author} {\bibinfo {author} {\bibfnamefont {D.~P.}\ \bibnamefont
  {Arovas}}, \bibinfo {author} {\bibfnamefont {E.}~\bibnamefont {Berg}},
  \bibinfo {author} {\bibfnamefont {S.~A.}\ \bibnamefont {Kivelson}},\ and\
  \bibinfo {author} {\bibfnamefont {S.}~\bibnamefont {Raghu}},\ }\href@noop {}
  {\bibfield  {journal} {\bibinfo  {journal} {Annual review of condensed matter
  physics}\ }\textbf {\bibinfo {volume} {13}},\ \bibinfo {pages} {239}
  (\bibinfo {year} {2022})}\BibitemShut {NoStop}%
\bibitem [{\citenamefont {Auerbach}(2012)}]{auerbach2012interacting}%
  \BibitemOpen
  \bibfield  {author} {\bibinfo {author} {\bibfnamefont {A.}~\bibnamefont
  {Auerbach}},\ }\href@noop {} {\emph {\bibinfo {title} {Interacting electrons
  and quantum magnetism}}}\ (\bibinfo  {publisher} {Springer Science \&
  Business Media},\ \bibinfo {year} {2012})\BibitemShut {NoStop}%
\bibitem [{\citenamefont {Lieb}\ and\ \citenamefont {Wu}(1968)}]{Lieb1968}%
  \BibitemOpen
  \bibfield  {author} {\bibinfo {author} {\bibfnamefont {E.~H.}\ \bibnamefont
  {Lieb}}\ and\ \bibinfo {author} {\bibfnamefont {F.~Y.}\ \bibnamefont {Wu}},\
  }\href {https://doi.org/10.1103/PhysRevLett.20.1445} {\bibfield  {journal}
  {\bibinfo  {journal} {Phys. Rev. Lett.}\ }\textbf {\bibinfo {volume} {20}},\
  \bibinfo {pages} {1445} (\bibinfo {year} {1968})}\BibitemShut {NoStop}%
\bibitem [{\citenamefont {Giamarchi}(2003)}]{Giamarchi2003}%
  \BibitemOpen
  \bibfield  {author} {\bibinfo {author} {\bibfnamefont {T.}~\bibnamefont
  {Giamarchi}},\ }\href@noop {} {\emph {\bibinfo {title} {Quantum physics in
  one dimension}}},\ Vol.\ \bibinfo {volume} {121}\ (\bibinfo  {publisher}
  {Clarendon press},\ \bibinfo {year} {2003})\BibitemShut {NoStop}%
\bibitem [{\citenamefont {Essler}\ \emph {et~al.}(2005)\citenamefont {Essler},
  \citenamefont {Frahm}, \citenamefont {G{\"o}hmann}, \citenamefont
  {Kl{\"u}mper},\ and\ \citenamefont {Korepin}}]{essler2005one}%
  \BibitemOpen
  \bibfield  {author} {\bibinfo {author} {\bibfnamefont {F.~H.}\ \bibnamefont
  {Essler}}, \bibinfo {author} {\bibfnamefont {H.}~\bibnamefont {Frahm}},
  \bibinfo {author} {\bibfnamefont {F.}~\bibnamefont {G{\"o}hmann}}, \bibinfo
  {author} {\bibfnamefont {A.}~\bibnamefont {Kl{\"u}mper}},\ and\ \bibinfo
  {author} {\bibfnamefont {V.~E.}\ \bibnamefont {Korepin}},\ }\href@noop {}
  {\emph {\bibinfo {title} {The one-dimensional Hubbard model}}}\ (\bibinfo
  {publisher} {Cambridge University Press},\ \bibinfo {year}
  {2005})\BibitemShut {NoStop}%
\bibitem [{\citenamefont {Hirsch}(1985)}]{hirsch1985two}%
  \BibitemOpen
  \bibfield  {author} {\bibinfo {author} {\bibfnamefont {J.~E.}\ \bibnamefont
  {Hirsch}},\ }\href@noop {} {\bibfield  {journal} {\bibinfo  {journal}
  {Physical Review B}\ }\textbf {\bibinfo {volume} {31}},\ \bibinfo {pages}
  {4403} (\bibinfo {year} {1985})}\BibitemShut {NoStop}%
\bibitem [{\citenamefont {White}\ \emph {et~al.}(1989)\citenamefont {White},
  \citenamefont {Scalapino}, \citenamefont {Sugar}, \citenamefont {Loh},
  \citenamefont {Gubernatis},\ and\ \citenamefont
  {Scalettar}}]{white1989numerical}%
  \BibitemOpen
  \bibfield  {author} {\bibinfo {author} {\bibfnamefont {S.~R.}\ \bibnamefont
  {White}}, \bibinfo {author} {\bibfnamefont {D.~J.}\ \bibnamefont
  {Scalapino}}, \bibinfo {author} {\bibfnamefont {R.~L.}\ \bibnamefont
  {Sugar}}, \bibinfo {author} {\bibfnamefont {E.~Y.}\ \bibnamefont {Loh}},
  \bibinfo {author} {\bibfnamefont {J.~E.}\ \bibnamefont {Gubernatis}},\ and\
  \bibinfo {author} {\bibfnamefont {R.~T.}\ \bibnamefont {Scalettar}},\ }\href
  {https://doi.org/10.1103/PhysRevB.40.506} {\bibfield  {journal} {\bibinfo
  {journal} {Phys. Rev. B}\ }\textbf {\bibinfo {volume} {40}},\ \bibinfo
  {pages} {506} (\bibinfo {year} {1989})}\BibitemShut {NoStop}%
\bibitem [{\citenamefont {LeBlanc}\ \emph {et~al.}(2015)\citenamefont
  {LeBlanc}, \citenamefont {Antipov}, \citenamefont {Becca}, \citenamefont
  {Bulik}, \citenamefont {Chan}, \citenamefont {Chung}, \citenamefont {Deng},
  \citenamefont {Ferrero}, \citenamefont {Henderson}, \citenamefont
  {Jim\'enez-Hoyos}, \citenamefont {Kozik}, \citenamefont {Liu}, \citenamefont
  {Millis}, \citenamefont {Prokof'ev}, \citenamefont {Qin}, \citenamefont
  {Scuseria}, \citenamefont {Shi}, \citenamefont {Svistunov}, \citenamefont
  {Tocchio}, \citenamefont {Tupitsyn}, \citenamefont {White}, \citenamefont
  {Zhang}, \citenamefont {Zheng}, \citenamefont {Zhu},\ and\ \citenamefont
  {Gull}}]{leblanc2015solutions}%
  \BibitemOpen
  \bibfield  {author} {\bibinfo {author} {\bibfnamefont {J.~P.~F.}\
  \bibnamefont {LeBlanc}}, \bibinfo {author} {\bibfnamefont {A.~E.}\
  \bibnamefont {Antipov}}, \bibinfo {author} {\bibfnamefont {F.}~\bibnamefont
  {Becca}}, \bibinfo {author} {\bibfnamefont {I.~W.}\ \bibnamefont {Bulik}},
  \bibinfo {author} {\bibfnamefont {G.~K.-L.}\ \bibnamefont {Chan}}, \bibinfo
  {author} {\bibfnamefont {C.-M.}\ \bibnamefont {Chung}}, \bibinfo {author}
  {\bibfnamefont {Y.}~\bibnamefont {Deng}}, \bibinfo {author} {\bibfnamefont
  {M.}~\bibnamefont {Ferrero}}, \bibinfo {author} {\bibfnamefont {T.~M.}\
  \bibnamefont {Henderson}}, \bibinfo {author} {\bibfnamefont {C.~A.}\
  \bibnamefont {Jim\'enez-Hoyos}}, \bibinfo {author} {\bibfnamefont
  {E.}~\bibnamefont {Kozik}}, \bibinfo {author} {\bibfnamefont {X.-W.}\
  \bibnamefont {Liu}}, \bibinfo {author} {\bibfnamefont {A.~J.}\ \bibnamefont
  {Millis}}, \bibinfo {author} {\bibfnamefont {N.~V.}\ \bibnamefont
  {Prokof'ev}}, \bibinfo {author} {\bibfnamefont {M.}~\bibnamefont {Qin}},
  \bibinfo {author} {\bibfnamefont {G.~E.}\ \bibnamefont {Scuseria}}, \bibinfo
  {author} {\bibfnamefont {H.}~\bibnamefont {Shi}}, \bibinfo {author}
  {\bibfnamefont {B.~V.}\ \bibnamefont {Svistunov}}, \bibinfo {author}
  {\bibfnamefont {L.~F.}\ \bibnamefont {Tocchio}}, \bibinfo {author}
  {\bibfnamefont {I.~S.}\ \bibnamefont {Tupitsyn}}, \bibinfo {author}
  {\bibfnamefont {S.~R.}\ \bibnamefont {White}}, \bibinfo {author}
  {\bibfnamefont {S.}~\bibnamefont {Zhang}}, \bibinfo {author} {\bibfnamefont
  {B.-X.}\ \bibnamefont {Zheng}}, \bibinfo {author} {\bibfnamefont
  {Z.}~\bibnamefont {Zhu}},\ and\ \bibinfo {author} {\bibfnamefont
  {E.}~\bibnamefont {Gull}} (\bibinfo {collaboration} {Simons Collaboration on
  the Many-Electron Problem}),\ }\href
  {https://doi.org/10.1103/PhysRevX.5.041041} {\bibfield  {journal} {\bibinfo
  {journal} {Phys. Rev. X}\ }\textbf {\bibinfo {volume} {5}},\ \bibinfo {pages}
  {041041} (\bibinfo {year} {2015})}\BibitemShut {NoStop}%
\bibitem [{\citenamefont {Vannimenus}\ and\ \citenamefont
  {Toulouse}(1977)}]{Vannimenus1977}%
  \BibitemOpen
  \bibfield  {author} {\bibinfo {author} {\bibfnamefont {J.}~\bibnamefont
  {Vannimenus}}\ and\ \bibinfo {author} {\bibfnamefont {G.}~\bibnamefont
  {Toulouse}},\ }\href {https://doi.org/doi:10.1088/0022-3719/10/18/008}
  {\bibfield  {journal} {\bibinfo  {journal} {J. Phys. C}\ }\textbf {\bibinfo
  {volume} {10}},\ \bibinfo {pages} {L537} (\bibinfo {year}
  {1977})}\BibitemShut {NoStop}%
\bibitem [{\citenamefont {Toulouse}\ \emph {et~al.}(1987)\citenamefont
  {Toulouse} \emph {et~al.}}]{Toulouse1987}%
  \BibitemOpen
  \bibfield  {author} {\bibinfo {author} {\bibfnamefont {G.}~\bibnamefont
  {Toulouse}} \emph {et~al.},\ }\href@noop {} {\bibfield  {journal} {\bibinfo
  {journal} {Spin Glass Theory and Beyond: An Introduction to the Replica
  Method and Its Applications}\ }\textbf {\bibinfo {volume} {9}},\ \bibinfo
  {pages} {99} (\bibinfo {year} {1987})}\BibitemShut {NoStop}%
\bibitem [{\citenamefont {Diep}(2004)}]{BookDiep}%
  \BibitemOpen
  \bibfield  {author} {\bibinfo {author} {\bibfnamefont {H.~T.}\ \bibnamefont
  {Diep}},\ }\href@noop {} {\emph {\bibinfo {title} {Frustrated spin
  systems}}}\ (\bibinfo  {publisher} {World Scientific},\ \bibinfo {year}
  {2004})\BibitemShut {NoStop}%
\bibitem [{\citenamefont {Lacroix}\ \emph {et~al.}(2011)\citenamefont
  {Lacroix}, \citenamefont {Mendels},\ and\ \citenamefont
  {Mila}}]{BookLacroix}%
  \BibitemOpen
  \bibfield  {author} {\bibinfo {author} {\bibfnamefont {C.}~\bibnamefont
  {Lacroix}}, \bibinfo {author} {\bibfnamefont {P.}~\bibnamefont {Mendels}},\
  and\ \bibinfo {author} {\bibfnamefont {F.}~\bibnamefont {Mila}},\ }\href@noop
  {} {\emph {\bibinfo {title} {Introduction to Frustrated Magnetism: Materials,
  Experiments, Theory}}},\ Vol.\ \bibinfo {volume} {164}\ (\bibinfo
  {publisher} {Springer},\ \bibinfo {year} {2011})\BibitemShut {NoStop}%
\bibitem [{\citenamefont {Anderson}(1973)}]{Anderson73}%
  \BibitemOpen
  \bibfield  {author} {\bibinfo {author} {\bibfnamefont {P.~W.}\ \bibnamefont
  {Anderson}},\ }\href {https://doi.org/10.1016/0025-5408(73)90167-0}
  {\bibfield  {journal} {\bibinfo  {journal} {Mater. Res. Bull.}\ }\textbf
  {\bibinfo {volume} {8}},\ \bibinfo {pages} {153} (\bibinfo {year}
  {1973})}\BibitemShut {NoStop}%
\bibitem [{\citenamefont {Anderson}(1987)}]{Anderson87}%
  \BibitemOpen
  \bibfield  {author} {\bibinfo {author} {\bibfnamefont {P.~W.}\ \bibnamefont
  {Anderson}},\ }\href {https://doi.org/10.1126/science.235.4793.1196}
  {\bibfield  {journal} {\bibinfo  {journal} {Science}\ }\textbf {\bibinfo
  {volume} {235}},\ \bibinfo {pages} {1196} (\bibinfo {year}
  {1987})}\BibitemShut {NoStop}%
\bibitem [{\citenamefont {Lee}(2008)}]{Lee08}%
  \BibitemOpen
  \bibfield  {author} {\bibinfo {author} {\bibfnamefont {P.~A.}\ \bibnamefont
  {Lee}},\ }\href {https://doi.org/10.1126/science.1163196} {\bibfield
  {journal} {\bibinfo  {journal} {Science}\ }\textbf {\bibinfo {volume}
  {321}},\ \bibinfo {pages} {1306} (\bibinfo {year} {2008})}\BibitemShut
  {NoStop}%
\bibitem [{\citenamefont {Balents}(2010)}]{Balents10}%
  \BibitemOpen
  \bibfield  {author} {\bibinfo {author} {\bibfnamefont {L.}~\bibnamefont
  {Balents}},\ }\href {http://dx.doi.org/10.1038/nature08917} {\bibfield
  {journal} {\bibinfo  {journal} {Nature}\ }\textbf {\bibinfo {volume} {464}},\
  \bibinfo {pages} {199} (\bibinfo {year} {2010})}\BibitemShut {NoStop}%
\bibitem [{\citenamefont {Savary}\ and\ \citenamefont
  {Balents}(2016)}]{Savary2016}%
  \BibitemOpen
  \bibfield  {author} {\bibinfo {author} {\bibfnamefont {L.}~\bibnamefont
  {Savary}}\ and\ \bibinfo {author} {\bibfnamefont {L.}~\bibnamefont
  {Balents}},\ }\href {https://doi.org/10.1088/0034-4885/80/1/016502}
  {\bibfield  {journal} {\bibinfo  {journal} {Rep. Prog. Phys.}\ }\textbf
  {\bibinfo {volume} {80}},\ \bibinfo {pages} {016502} (\bibinfo {year}
  {2016})}\BibitemShut {NoStop}%
\bibitem [{\citenamefont {Zhou}\ \emph {et~al.}(2017)\citenamefont {Zhou},
  \citenamefont {Kanoda},\ and\ \citenamefont {Ng}}]{QSLRMP}%
  \BibitemOpen
  \bibfield  {author} {\bibinfo {author} {\bibfnamefont {Y.}~\bibnamefont
  {Zhou}}, \bibinfo {author} {\bibfnamefont {K.}~\bibnamefont {Kanoda}},\ and\
  \bibinfo {author} {\bibfnamefont {T.-K.}\ \bibnamefont {Ng}},\ }\href
  {https://doi.org/10.1103/RevModPhys.89.025003} {\bibfield  {journal}
  {\bibinfo  {journal} {Rev. Mod. Phys.}\ }\textbf {\bibinfo {volume} {89}},\
  \bibinfo {pages} {025003} (\bibinfo {year} {2017})}\BibitemShut {NoStop}%
\bibitem [{\citenamefont {Knolle}\ and\ \citenamefont
  {Moessner}(2019)}]{Knolle2019}%
  \BibitemOpen
  \bibfield  {author} {\bibinfo {author} {\bibfnamefont {J.}~\bibnamefont
  {Knolle}}\ and\ \bibinfo {author} {\bibfnamefont {R.}~\bibnamefont
  {Moessner}},\ }\href
  {https://doi.org/10.1146/annurev-conmatphys-031218-013401} {\bibfield
  {journal} {\bibinfo  {journal} {Annu. Rev. Condens. Matter Phys.}\ }\textbf
  {\bibinfo {volume} {10}},\ \bibinfo {pages} {451} (\bibinfo {year}
  {2019})}\BibitemShut {NoStop}%
\bibitem [{\citenamefont {Broholm}\ \emph {et~al.}(2020)\citenamefont
  {Broholm}, \citenamefont {Cava}, \citenamefont {Kivelson}, \citenamefont
  {Nocera}, \citenamefont {Norman},\ and\ \citenamefont
  {Senthil}}]{Broholm2020}%
  \BibitemOpen
  \bibfield  {author} {\bibinfo {author} {\bibfnamefont {C.}~\bibnamefont
  {Broholm}}, \bibinfo {author} {\bibfnamefont {R.~J.}\ \bibnamefont {Cava}},
  \bibinfo {author} {\bibfnamefont {S.~A.}\ \bibnamefont {Kivelson}}, \bibinfo
  {author} {\bibfnamefont {D.~G.}\ \bibnamefont {Nocera}}, \bibinfo {author}
  {\bibfnamefont {M.~R.}\ \bibnamefont {Norman}},\ and\ \bibinfo {author}
  {\bibfnamefont {T.}~\bibnamefont {Senthil}},\ }\href
  {https://science.sciencemag.org/content/367/6475/eaay0668} {\bibfield
  {journal} {\bibinfo  {journal} {Science}\ }\textbf {\bibinfo {volume} {367}}
  (\bibinfo {year} {2020})}\BibitemShut {NoStop}%
\bibitem [{\citenamefont {Li}\ \emph {et~al.}(2020)\citenamefont {Li},
  \citenamefont {Gegenwart},\ and\ \citenamefont {Tsirlin}}]{li2020spin}%
  \BibitemOpen
  \bibfield  {author} {\bibinfo {author} {\bibfnamefont {Y.}~\bibnamefont
  {Li}}, \bibinfo {author} {\bibfnamefont {P.}~\bibnamefont {Gegenwart}},\ and\
  \bibinfo {author} {\bibfnamefont {A.~A.}\ \bibnamefont {Tsirlin}},\
  }\href@noop {} {\bibfield  {journal} {\bibinfo  {journal} {Journal of
  Physics: Condensed Matter}\ }\textbf {\bibinfo {volume} {32}},\ \bibinfo
  {pages} {224004} (\bibinfo {year} {2020})}\BibitemShut {NoStop}%
\bibitem [{\citenamefont {Kanoda}\ and\ \citenamefont
  {Kato}(2011)}]{Kanoda2011}%
  \BibitemOpen
  \bibfield  {author} {\bibinfo {author} {\bibfnamefont {K.}~\bibnamefont
  {Kanoda}}\ and\ \bibinfo {author} {\bibfnamefont {R.}~\bibnamefont {Kato}},\
  }\href {https://doi.org/10.1146/annurev-conmatphys-062910-140521} {\bibfield
  {journal} {\bibinfo  {journal} {Annual Review of Condensed Matter Physics}\
  }\textbf {\bibinfo {volume} {2}},\ \bibinfo {pages} {167} (\bibinfo {year}
  {2011})}\BibitemShut {NoStop}%
\bibitem [{\citenamefont {Powell}\ and\ \citenamefont
  {McKenzie}(2011)}]{Powell2011}%
  \BibitemOpen
  \bibfield  {author} {\bibinfo {author} {\bibfnamefont {B.~J.}\ \bibnamefont
  {Powell}}\ and\ \bibinfo {author} {\bibfnamefont {R.~H.}\ \bibnamefont
  {McKenzie}},\ }\href {https://doi.org/10.1088/0034-4885/74/5/056501}
  {\bibfield  {journal} {\bibinfo  {journal} {Reports on Progress in Physics}\
  }\textbf {\bibinfo {volume} {74}},\ \bibinfo {pages} {056501} (\bibinfo
  {year} {2011})}\BibitemShut {NoStop}%
\bibitem [{\citenamefont {Li}\ \emph {et~al.}(2015)\citenamefont {Li},
  \citenamefont {Liao}, \citenamefont {Zhang}, \citenamefont {Li},
  \citenamefont {Jin}, \citenamefont {Ling}, \citenamefont {Zhang},
  \citenamefont {Zou}, \citenamefont {Pi}, \citenamefont {Yang} \emph
  {et~al.}}]{Li2015}%
  \BibitemOpen
  \bibfield  {author} {\bibinfo {author} {\bibfnamefont {Y.}~\bibnamefont
  {Li}}, \bibinfo {author} {\bibfnamefont {H.}~\bibnamefont {Liao}}, \bibinfo
  {author} {\bibfnamefont {Z.}~\bibnamefont {Zhang}}, \bibinfo {author}
  {\bibfnamefont {S.}~\bibnamefont {Li}}, \bibinfo {author} {\bibfnamefont
  {F.}~\bibnamefont {Jin}}, \bibinfo {author} {\bibfnamefont {L.}~\bibnamefont
  {Ling}}, \bibinfo {author} {\bibfnamefont {L.}~\bibnamefont {Zhang}},
  \bibinfo {author} {\bibfnamefont {Y.}~\bibnamefont {Zou}}, \bibinfo {author}
  {\bibfnamefont {L.}~\bibnamefont {Pi}}, \bibinfo {author} {\bibfnamefont
  {Z.}~\bibnamefont {Yang}}, \emph {et~al.},\ }\href
  {https://doi.org/10.1038/srep16419} {\bibfield  {journal} {\bibinfo
  {journal} {Scientific reports}\ }\textbf {\bibinfo {volume} {5}},\ \bibinfo
  {pages} {1} (\bibinfo {year} {2015})}\BibitemShut {NoStop}%
\bibitem [{\citenamefont {Shen}\ \emph {et~al.}(2016)\citenamefont {Shen},
  \citenamefont {Li}, \citenamefont {Wo}, \citenamefont {Li}, \citenamefont
  {Shen}, \citenamefont {Pan}, \citenamefont {Wang}, \citenamefont {Walker},
  \citenamefont {Steffens}, \citenamefont {Boehm} \emph {et~al.}}]{Shen2016}%
  \BibitemOpen
  \bibfield  {author} {\bibinfo {author} {\bibfnamefont {Y.}~\bibnamefont
  {Shen}}, \bibinfo {author} {\bibfnamefont {Y.-D.}\ \bibnamefont {Li}},
  \bibinfo {author} {\bibfnamefont {H.}~\bibnamefont {Wo}}, \bibinfo {author}
  {\bibfnamefont {Y.}~\bibnamefont {Li}}, \bibinfo {author} {\bibfnamefont
  {S.}~\bibnamefont {Shen}}, \bibinfo {author} {\bibfnamefont {B.}~\bibnamefont
  {Pan}}, \bibinfo {author} {\bibfnamefont {Q.}~\bibnamefont {Wang}}, \bibinfo
  {author} {\bibfnamefont {H.}~\bibnamefont {Walker}}, \bibinfo {author}
  {\bibfnamefont {P.}~\bibnamefont {Steffens}}, \bibinfo {author}
  {\bibfnamefont {M.}~\bibnamefont {Boehm}}, \emph {et~al.},\ }\href
  {https://doi.org/10.1038/nature20614} {\bibfield  {journal} {\bibinfo
  {journal} {Nature}\ }\textbf {\bibinfo {volume} {540}},\ \bibinfo {pages}
  {559} (\bibinfo {year} {2016})}\BibitemShut {NoStop}%
\bibitem [{\citenamefont {Ding}\ \emph {et~al.}(2019)\citenamefont {Ding},
  \citenamefont {Manuel}, \citenamefont {Bachus}, \citenamefont {Gru{\ss}ler},
  \citenamefont {Gegenwart}, \citenamefont {Singleton}, \citenamefont
  {Johnson}, \citenamefont {Walker}, \citenamefont {Adroja}, \citenamefont
  {Hillier} \emph {et~al.}}]{ding2019gapless}%
  \BibitemOpen
  \bibfield  {author} {\bibinfo {author} {\bibfnamefont {L.}~\bibnamefont
  {Ding}}, \bibinfo {author} {\bibfnamefont {P.}~\bibnamefont {Manuel}},
  \bibinfo {author} {\bibfnamefont {S.}~\bibnamefont {Bachus}}, \bibinfo
  {author} {\bibfnamefont {F.}~\bibnamefont {Gru{\ss}ler}}, \bibinfo {author}
  {\bibfnamefont {P.}~\bibnamefont {Gegenwart}}, \bibinfo {author}
  {\bibfnamefont {J.}~\bibnamefont {Singleton}}, \bibinfo {author}
  {\bibfnamefont {R.~D.}\ \bibnamefont {Johnson}}, \bibinfo {author}
  {\bibfnamefont {H.~C.}\ \bibnamefont {Walker}}, \bibinfo {author}
  {\bibfnamefont {D.~T.}\ \bibnamefont {Adroja}}, \bibinfo {author}
  {\bibfnamefont {A.~D.}\ \bibnamefont {Hillier}}, \emph {et~al.},\ }\href@noop
  {} {\bibfield  {journal} {\bibinfo  {journal} {Physical Review B}\ }\textbf
  {\bibinfo {volume} {100}},\ \bibinfo {pages} {144432} (\bibinfo {year}
  {2019})}\BibitemShut {NoStop}%
\bibitem [{\citenamefont {Cao}\ \emph {et~al.}(2018{\natexlab{a}})\citenamefont
  {Cao}, \citenamefont {Fatemi}, \citenamefont {Demir}, \citenamefont {Fang},
  \citenamefont {Tomarken}, \citenamefont {Luo}, \citenamefont
  {Sanchez-Yamagishi}, \citenamefont {Watanabe}, \citenamefont {Taniguchi},
  \citenamefont {Kaxiras} \emph {et~al.}}]{Cao2018}%
  \BibitemOpen
  \bibfield  {author} {\bibinfo {author} {\bibfnamefont {Y.}~\bibnamefont
  {Cao}}, \bibinfo {author} {\bibfnamefont {V.}~\bibnamefont {Fatemi}},
  \bibinfo {author} {\bibfnamefont {A.}~\bibnamefont {Demir}}, \bibinfo
  {author} {\bibfnamefont {S.}~\bibnamefont {Fang}}, \bibinfo {author}
  {\bibfnamefont {S.~L.}\ \bibnamefont {Tomarken}}, \bibinfo {author}
  {\bibfnamefont {J.~Y.}\ \bibnamefont {Luo}}, \bibinfo {author} {\bibfnamefont
  {J.~D.}\ \bibnamefont {Sanchez-Yamagishi}}, \bibinfo {author} {\bibfnamefont
  {K.}~\bibnamefont {Watanabe}}, \bibinfo {author} {\bibfnamefont
  {T.}~\bibnamefont {Taniguchi}}, \bibinfo {author} {\bibfnamefont
  {E.}~\bibnamefont {Kaxiras}}, \emph {et~al.},\ }\href@noop {} {\bibfield
  {journal} {\bibinfo  {journal} {Nature}\ }\textbf {\bibinfo {volume} {556}},\
  \bibinfo {pages} {80} (\bibinfo {year} {2018}{\natexlab{a}})}\BibitemShut
  {NoStop}%
\bibitem [{\citenamefont {Cao}\ \emph {et~al.}(2018{\natexlab{b}})\citenamefont
  {Cao}, \citenamefont {Fatemi}, \citenamefont {Fang}, \citenamefont
  {Watanabe}, \citenamefont {Taniguchi}, \citenamefont {Kaxiras},\ and\
  \citenamefont {Jarillo-Herrero}}]{Cao2018_2}%
  \BibitemOpen
  \bibfield  {author} {\bibinfo {author} {\bibfnamefont {Y.}~\bibnamefont
  {Cao}}, \bibinfo {author} {\bibfnamefont {V.}~\bibnamefont {Fatemi}},
  \bibinfo {author} {\bibfnamefont {S.}~\bibnamefont {Fang}}, \bibinfo {author}
  {\bibfnamefont {K.}~\bibnamefont {Watanabe}}, \bibinfo {author}
  {\bibfnamefont {T.}~\bibnamefont {Taniguchi}}, \bibinfo {author}
  {\bibfnamefont {E.}~\bibnamefont {Kaxiras}},\ and\ \bibinfo {author}
  {\bibfnamefont {P.}~\bibnamefont {Jarillo-Herrero}},\ }\href@noop {}
  {\bibfield  {journal} {\bibinfo  {journal} {Nature}\ }\textbf {\bibinfo
  {volume} {556}},\ \bibinfo {pages} {43} (\bibinfo {year}
  {2018}{\natexlab{b}})}\BibitemShut {NoStop}%
\bibitem [{\citenamefont {Yankowitz}\ \emph {et~al.}(2019)\citenamefont
  {Yankowitz}, \citenamefont {Chen}, \citenamefont {Polshyn}, \citenamefont
  {Zhang}, \citenamefont {Watanabe}, \citenamefont {Taniguchi}, \citenamefont
  {Graf}, \citenamefont {Young},\ and\ \citenamefont {Dean}}]{Yankowitz2019}%
  \BibitemOpen
  \bibfield  {author} {\bibinfo {author} {\bibfnamefont {M.}~\bibnamefont
  {Yankowitz}}, \bibinfo {author} {\bibfnamefont {S.}~\bibnamefont {Chen}},
  \bibinfo {author} {\bibfnamefont {H.}~\bibnamefont {Polshyn}}, \bibinfo
  {author} {\bibfnamefont {Y.}~\bibnamefont {Zhang}}, \bibinfo {author}
  {\bibfnamefont {K.}~\bibnamefont {Watanabe}}, \bibinfo {author}
  {\bibfnamefont {T.}~\bibnamefont {Taniguchi}}, \bibinfo {author}
  {\bibfnamefont {D.}~\bibnamefont {Graf}}, \bibinfo {author} {\bibfnamefont
  {A.~F.}\ \bibnamefont {Young}},\ and\ \bibinfo {author} {\bibfnamefont
  {C.~R.}\ \bibnamefont {Dean}},\ }\href@noop {} {\bibfield  {journal}
  {\bibinfo  {journal} {Science}\ }\textbf {\bibinfo {volume} {363}},\ \bibinfo
  {pages} {1059} (\bibinfo {year} {2019})}\BibitemShut {NoStop}%
\bibitem [{\citenamefont {Wu}\ \emph {et~al.}(2018)\citenamefont {Wu},
  \citenamefont {Lovorn}, \citenamefont {Tutuc},\ and\ \citenamefont
  {MacDonald}}]{Wu2018}%
  \BibitemOpen
  \bibfield  {author} {\bibinfo {author} {\bibfnamefont {F.}~\bibnamefont
  {Wu}}, \bibinfo {author} {\bibfnamefont {T.}~\bibnamefont {Lovorn}}, \bibinfo
  {author} {\bibfnamefont {E.}~\bibnamefont {Tutuc}},\ and\ \bibinfo {author}
  {\bibfnamefont {A.~H.}\ \bibnamefont {MacDonald}},\ }\href
  {https://doi.org/10.1103/PhysRevLett.121.026402} {\bibfield  {journal}
  {\bibinfo  {journal} {Phys. Rev. Lett.}\ }\textbf {\bibinfo {volume} {121}},\
  \bibinfo {pages} {026402} (\bibinfo {year} {2018})}\BibitemShut {NoStop}%
\bibitem [{\citenamefont {Wu}\ \emph {et~al.}(2019)\citenamefont {Wu},
  \citenamefont {Lovorn}, \citenamefont {Tutuc}, \citenamefont {Martin},\ and\
  \citenamefont {MacDonald}}]{Wu2019}%
  \BibitemOpen
  \bibfield  {author} {\bibinfo {author} {\bibfnamefont {F.}~\bibnamefont
  {Wu}}, \bibinfo {author} {\bibfnamefont {T.}~\bibnamefont {Lovorn}}, \bibinfo
  {author} {\bibfnamefont {E.}~\bibnamefont {Tutuc}}, \bibinfo {author}
  {\bibfnamefont {I.}~\bibnamefont {Martin}},\ and\ \bibinfo {author}
  {\bibfnamefont {A.~H.}\ \bibnamefont {MacDonald}},\ }\href
  {https://doi.org/10.1103/PhysRevLett.122.086402} {\bibfield  {journal}
  {\bibinfo  {journal} {Phys. Rev. Lett.}\ }\textbf {\bibinfo {volume} {122}},\
  \bibinfo {pages} {086402} (\bibinfo {year} {2019})}\BibitemShut {NoStop}%
\bibitem [{\citenamefont {Wang}\ \emph {et~al.}(2020)\citenamefont {Wang},
  \citenamefont {Shih}, \citenamefont {Ghiotto}, \citenamefont {Xian},
  \citenamefont {Rhodes}, \citenamefont {Tan}, \citenamefont {Claassen},
  \citenamefont {Kennes}, \citenamefont {Bai}, \citenamefont {Kim} \emph
  {et~al.}}]{Wang2020}%
  \BibitemOpen
  \bibfield  {author} {\bibinfo {author} {\bibfnamefont {L.}~\bibnamefont
  {Wang}}, \bibinfo {author} {\bibfnamefont {E.-M.}\ \bibnamefont {Shih}},
  \bibinfo {author} {\bibfnamefont {A.}~\bibnamefont {Ghiotto}}, \bibinfo
  {author} {\bibfnamefont {L.}~\bibnamefont {Xian}}, \bibinfo {author}
  {\bibfnamefont {D.~A.}\ \bibnamefont {Rhodes}}, \bibinfo {author}
  {\bibfnamefont {C.}~\bibnamefont {Tan}}, \bibinfo {author} {\bibfnamefont
  {M.}~\bibnamefont {Claassen}}, \bibinfo {author} {\bibfnamefont {D.~M.}\
  \bibnamefont {Kennes}}, \bibinfo {author} {\bibfnamefont {Y.}~\bibnamefont
  {Bai}}, \bibinfo {author} {\bibfnamefont {B.}~\bibnamefont {Kim}}, \emph
  {et~al.},\ }\href@noop {} {\bibfield  {journal} {\bibinfo  {journal} {Nature
  materials}\ }\textbf {\bibinfo {volume} {19}},\ \bibinfo {pages} {861}
  (\bibinfo {year} {2020})}\BibitemShut {NoStop}%
\bibitem [{\citenamefont {Kennes}\ \emph {et~al.}(2021)\citenamefont {Kennes},
  \citenamefont {Claassen}, \citenamefont {Xian}, \citenamefont {Georges},
  \citenamefont {Millis}, \citenamefont {Hone}, \citenamefont {Dean},
  \citenamefont {Basov}, \citenamefont {Pasupathy},\ and\ \citenamefont
  {Rubio}}]{Kennes2021}%
  \BibitemOpen
  \bibfield  {author} {\bibinfo {author} {\bibfnamefont {D.~M.}\ \bibnamefont
  {Kennes}}, \bibinfo {author} {\bibfnamefont {M.}~\bibnamefont {Claassen}},
  \bibinfo {author} {\bibfnamefont {L.}~\bibnamefont {Xian}}, \bibinfo {author}
  {\bibfnamefont {A.}~\bibnamefont {Georges}}, \bibinfo {author} {\bibfnamefont
  {A.~J.}\ \bibnamefont {Millis}}, \bibinfo {author} {\bibfnamefont
  {J.}~\bibnamefont {Hone}}, \bibinfo {author} {\bibfnamefont {C.~R.}\
  \bibnamefont {Dean}}, \bibinfo {author} {\bibfnamefont {D.}~\bibnamefont
  {Basov}}, \bibinfo {author} {\bibfnamefont {A.~N.}\ \bibnamefont
  {Pasupathy}},\ and\ \bibinfo {author} {\bibfnamefont {A.}~\bibnamefont
  {Rubio}},\ }\href@noop {} {\bibfield  {journal} {\bibinfo  {journal} {Nature
  Physics}\ }\textbf {\bibinfo {volume} {17}},\ \bibinfo {pages} {155}
  (\bibinfo {year} {2021})}\BibitemShut {NoStop}%
\bibitem [{\citenamefont {Kuhlenkamp}\ \emph {et~al.}(2022)\citenamefont
  {Kuhlenkamp}, \citenamefont {Kadow}, \citenamefont {Imamoglu},\ and\
  \citenamefont {Knap}}]{arxiv:2209.05506}%
  \BibitemOpen
  \bibfield  {author} {\bibinfo {author} {\bibfnamefont {C.}~\bibnamefont
  {Kuhlenkamp}}, \bibinfo {author} {\bibfnamefont {W.}~\bibnamefont {Kadow}},
  \bibinfo {author} {\bibfnamefont {A.}~\bibnamefont {Imamoglu}},\ and\
  \bibinfo {author} {\bibfnamefont {M.}~\bibnamefont {Knap}},\ }\href
  {https://doi.org/10.48550/ARXIV.2209.05506} {\bibinfo {title} {Tunable
  topological order of pseudo spins in semiconductor heterostructures}}
  (\bibinfo {year} {2022})\BibitemShut {NoStop}%
\bibitem [{\citenamefont {Ni}\ \emph {et~al.}(2019)\citenamefont {Ni},
  \citenamefont {Wang}, \citenamefont {Jiang}, \citenamefont {Chen},
  \citenamefont {Du}, \citenamefont {Sun}, \citenamefont {Goldflam},
  \citenamefont {Frenzel}, \citenamefont {Xie}, \citenamefont {Fogler} \emph
  {et~al.}}]{Ni2019}%
  \BibitemOpen
  \bibfield  {author} {\bibinfo {author} {\bibfnamefont {G.}~\bibnamefont
  {Ni}}, \bibinfo {author} {\bibfnamefont {H.}~\bibnamefont {Wang}}, \bibinfo
  {author} {\bibfnamefont {B.-Y.}\ \bibnamefont {Jiang}}, \bibinfo {author}
  {\bibfnamefont {L.}~\bibnamefont {Chen}}, \bibinfo {author} {\bibfnamefont
  {Y.}~\bibnamefont {Du}}, \bibinfo {author} {\bibfnamefont {Z.}~\bibnamefont
  {Sun}}, \bibinfo {author} {\bibfnamefont {M.}~\bibnamefont {Goldflam}},
  \bibinfo {author} {\bibfnamefont {A.}~\bibnamefont {Frenzel}}, \bibinfo
  {author} {\bibfnamefont {X.}~\bibnamefont {Xie}}, \bibinfo {author}
  {\bibfnamefont {M.}~\bibnamefont {Fogler}}, \emph {et~al.},\ }\href@noop {}
  {\bibfield  {journal} {\bibinfo  {journal} {Nature communications}\ }\textbf
  {\bibinfo {volume} {10}},\ \bibinfo {pages} {1} (\bibinfo {year}
  {2019})}\BibitemShut {NoStop}%
\bibitem [{\citenamefont {Xian}\ \emph {et~al.}(2019)\citenamefont {Xian},
  \citenamefont {Kennes}, \citenamefont {Tancogne-Dejean}, \citenamefont
  {Altarelli},\ and\ \citenamefont {Rubio}}]{Xian2019}%
  \BibitemOpen
  \bibfield  {author} {\bibinfo {author} {\bibfnamefont {L.}~\bibnamefont
  {Xian}}, \bibinfo {author} {\bibfnamefont {D.~M.}\ \bibnamefont {Kennes}},
  \bibinfo {author} {\bibfnamefont {N.}~\bibnamefont {Tancogne-Dejean}},
  \bibinfo {author} {\bibfnamefont {M.}~\bibnamefont {Altarelli}},\ and\
  \bibinfo {author} {\bibfnamefont {A.}~\bibnamefont {Rubio}},\ }\href@noop {}
  {\bibfield  {journal} {\bibinfo  {journal} {Nano letters}\ }\textbf {\bibinfo
  {volume} {19}},\ \bibinfo {pages} {4934} (\bibinfo {year}
  {2019})}\BibitemShut {NoStop}%
\bibitem [{\citenamefont {Tang}\ \emph {et~al.}(2020)\citenamefont {Tang},
  \citenamefont {Li}, \citenamefont {Li}, \citenamefont {Xu}, \citenamefont
  {Liu}, \citenamefont {Barmak}, \citenamefont {Watanabe}, \citenamefont
  {Taniguchi}, \citenamefont {MacDonald}, \citenamefont {Shan} \emph
  {et~al.}}]{Tang2020}%
  \BibitemOpen
  \bibfield  {author} {\bibinfo {author} {\bibfnamefont {Y.}~\bibnamefont
  {Tang}}, \bibinfo {author} {\bibfnamefont {L.}~\bibnamefont {Li}}, \bibinfo
  {author} {\bibfnamefont {T.}~\bibnamefont {Li}}, \bibinfo {author}
  {\bibfnamefont {Y.}~\bibnamefont {Xu}}, \bibinfo {author} {\bibfnamefont
  {S.}~\bibnamefont {Liu}}, \bibinfo {author} {\bibfnamefont {K.}~\bibnamefont
  {Barmak}}, \bibinfo {author} {\bibfnamefont {K.}~\bibnamefont {Watanabe}},
  \bibinfo {author} {\bibfnamefont {T.}~\bibnamefont {Taniguchi}}, \bibinfo
  {author} {\bibfnamefont {A.~H.}\ \bibnamefont {MacDonald}}, \bibinfo {author}
  {\bibfnamefont {J.}~\bibnamefont {Shan}}, \emph {et~al.},\ }\href@noop {}
  {\bibfield  {journal} {\bibinfo  {journal} {Nature}\ }\textbf {\bibinfo
  {volume} {579}},\ \bibinfo {pages} {353} (\bibinfo {year}
  {2020})}\BibitemShut {NoStop}%
\bibitem [{\citenamefont {Regan}\ \emph {et~al.}(2020)\citenamefont {Regan},
  \citenamefont {Wang}, \citenamefont {Jin}, \citenamefont {Bakti~Utama},
  \citenamefont {Gao}, \citenamefont {Wei}, \citenamefont {Zhao}, \citenamefont
  {Zhao}, \citenamefont {Zhang}, \citenamefont {Yumigeta} \emph
  {et~al.}}]{Regan2020}%
  \BibitemOpen
  \bibfield  {author} {\bibinfo {author} {\bibfnamefont {E.~C.}\ \bibnamefont
  {Regan}}, \bibinfo {author} {\bibfnamefont {D.}~\bibnamefont {Wang}},
  \bibinfo {author} {\bibfnamefont {C.}~\bibnamefont {Jin}}, \bibinfo {author}
  {\bibfnamefont {M.~I.}\ \bibnamefont {Bakti~Utama}}, \bibinfo {author}
  {\bibfnamefont {B.}~\bibnamefont {Gao}}, \bibinfo {author} {\bibfnamefont
  {X.}~\bibnamefont {Wei}}, \bibinfo {author} {\bibfnamefont {S.}~\bibnamefont
  {Zhao}}, \bibinfo {author} {\bibfnamefont {W.}~\bibnamefont {Zhao}}, \bibinfo
  {author} {\bibfnamefont {Z.}~\bibnamefont {Zhang}}, \bibinfo {author}
  {\bibfnamefont {K.}~\bibnamefont {Yumigeta}}, \emph {et~al.},\ }\href@noop {}
  {\bibfield  {journal} {\bibinfo  {journal} {Nature}\ }\textbf {\bibinfo
  {volume} {579}},\ \bibinfo {pages} {359} (\bibinfo {year}
  {2020})}\BibitemShut {NoStop}%
\bibitem [{\citenamefont {An}\ \emph {et~al.}(2020)\citenamefont {An},
  \citenamefont {Cai}, \citenamefont {Pei}, \citenamefont {Huang},
  \citenamefont {Wu}, \citenamefont {Zhou}, \citenamefont {Lin}, \citenamefont
  {Ying}, \citenamefont {Ye}, \citenamefont {Feng} \emph {et~al.}}]{An2020}%
  \BibitemOpen
  \bibfield  {author} {\bibinfo {author} {\bibfnamefont {L.}~\bibnamefont
  {An}}, \bibinfo {author} {\bibfnamefont {X.}~\bibnamefont {Cai}}, \bibinfo
  {author} {\bibfnamefont {D.}~\bibnamefont {Pei}}, \bibinfo {author}
  {\bibfnamefont {M.}~\bibnamefont {Huang}}, \bibinfo {author} {\bibfnamefont
  {Z.}~\bibnamefont {Wu}}, \bibinfo {author} {\bibfnamefont {Z.}~\bibnamefont
  {Zhou}}, \bibinfo {author} {\bibfnamefont {J.}~\bibnamefont {Lin}}, \bibinfo
  {author} {\bibfnamefont {Z.}~\bibnamefont {Ying}}, \bibinfo {author}
  {\bibfnamefont {Z.}~\bibnamefont {Ye}}, \bibinfo {author} {\bibfnamefont
  {X.}~\bibnamefont {Feng}}, \emph {et~al.},\ }\href@noop {} {\bibfield
  {journal} {\bibinfo  {journal} {Nanoscale horizons}\ }\textbf {\bibinfo
  {volume} {5}},\ \bibinfo {pages} {1309} (\bibinfo {year} {2020})}\BibitemShut
  {NoStop}%
\bibitem [{\citenamefont {Huse}\ and\ \citenamefont {Elser}(1988)}]{Huse1988}%
  \BibitemOpen
  \bibfield  {author} {\bibinfo {author} {\bibfnamefont {D.~A.}\ \bibnamefont
  {Huse}}\ and\ \bibinfo {author} {\bibfnamefont {V.}~\bibnamefont {Elser}},\
  }\href {https://doi.org/10.1103/PhysRevLett.60.2531} {\bibfield  {journal}
  {\bibinfo  {journal} {Phys. Rev. Lett.}\ }\textbf {\bibinfo {volume} {60}},\
  \bibinfo {pages} {2531} (\bibinfo {year} {1988})}\BibitemShut {NoStop}%
\bibitem [{\citenamefont {Bernu}\ \emph {et~al.}(1994)\citenamefont {Bernu},
  \citenamefont {Lecheminant}, \citenamefont {Lhuillier},\ and\ \citenamefont
  {Pierre}}]{Bernu1994}%
  \BibitemOpen
  \bibfield  {author} {\bibinfo {author} {\bibfnamefont {B.}~\bibnamefont
  {Bernu}}, \bibinfo {author} {\bibfnamefont {P.}~\bibnamefont {Lecheminant}},
  \bibinfo {author} {\bibfnamefont {C.}~\bibnamefont {Lhuillier}},\ and\
  \bibinfo {author} {\bibfnamefont {L.}~\bibnamefont {Pierre}},\ }\href
  {https://doi.org/10.1103/PhysRevB.50.10048} {\bibfield  {journal} {\bibinfo
  {journal} {Phys. Rev. B}\ }\textbf {\bibinfo {volume} {50}},\ \bibinfo
  {pages} {10048} (\bibinfo {year} {1994})}\BibitemShut {NoStop}%
\bibitem [{\citenamefont {Capriotti}\ \emph {et~al.}(1999)\citenamefont
  {Capriotti}, \citenamefont {Trumper},\ and\ \citenamefont
  {Sorella}}]{Capriotti1999}%
  \BibitemOpen
  \bibfield  {author} {\bibinfo {author} {\bibfnamefont {L.}~\bibnamefont
  {Capriotti}}, \bibinfo {author} {\bibfnamefont {A.~E.}\ \bibnamefont
  {Trumper}},\ and\ \bibinfo {author} {\bibfnamefont {S.}~\bibnamefont
  {Sorella}},\ }\href {https://doi.org/10.1103/PhysRevLett.82.3899} {\bibfield
  {journal} {\bibinfo  {journal} {Phys. Rev. Lett.}\ }\textbf {\bibinfo
  {volume} {82}},\ \bibinfo {pages} {3899} (\bibinfo {year}
  {1999})}\BibitemShut {NoStop}%
\bibitem [{\citenamefont {White}\ and\ \citenamefont
  {Chernyshev}(2007)}]{White2007}%
  \BibitemOpen
  \bibfield  {author} {\bibinfo {author} {\bibfnamefont {S.~R.}\ \bibnamefont
  {White}}\ and\ \bibinfo {author} {\bibfnamefont {A.~L.}\ \bibnamefont
  {Chernyshev}},\ }\href {https://doi.org/10.1103/PhysRevLett.99.127004}
  {\bibfield  {journal} {\bibinfo  {journal} {Phys. Rev. Lett.}\ }\textbf
  {\bibinfo {volume} {99}},\ \bibinfo {pages} {127004} (\bibinfo {year}
  {2007})}\BibitemShut {NoStop}%
\bibitem [{\citenamefont {Li}\ \emph {et~al.}(2022)\citenamefont {Li},
  \citenamefont {Li}, \citenamefont {Zhao}, \citenamefont {Luo},\ and\
  \citenamefont {Xie}}]{LiQian2022}%
  \BibitemOpen
  \bibfield  {author} {\bibinfo {author} {\bibfnamefont {Q.}~\bibnamefont
  {Li}}, \bibinfo {author} {\bibfnamefont {H.}~\bibnamefont {Li}}, \bibinfo
  {author} {\bibfnamefont {J.}~\bibnamefont {Zhao}}, \bibinfo {author}
  {\bibfnamefont {H.-G.}\ \bibnamefont {Luo}},\ and\ \bibinfo {author}
  {\bibfnamefont {Z.~Y.}\ \bibnamefont {Xie}},\ }\href
  {https://doi.org/10.1103/PhysRevB.105.184418} {\bibfield  {journal} {\bibinfo
   {journal} {Phys. Rev. B}\ }\textbf {\bibinfo {volume} {105}},\ \bibinfo
  {pages} {184418} (\bibinfo {year} {2022})}\BibitemShut {NoStop}%
\bibitem [{\citenamefont {Singh}(2005)}]{PhysRevB.71.214406}%
  \BibitemOpen
  \bibfield  {author} {\bibinfo {author} {\bibfnamefont {A.}~\bibnamefont
  {Singh}},\ }\href {https://doi.org/10.1103/PhysRevB.71.214406} {\bibfield
  {journal} {\bibinfo  {journal} {Phys. Rev. B}\ }\textbf {\bibinfo {volume}
  {71}},\ \bibinfo {pages} {214406} (\bibinfo {year} {2005})}\BibitemShut
  {NoStop}%
\bibitem [{\citenamefont {Sahebsara}\ and\ \citenamefont
  {S\'en\'echal}(2008)}]{Sahebsara2008}%
  \BibitemOpen
  \bibfield  {author} {\bibinfo {author} {\bibfnamefont {P.}~\bibnamefont
  {Sahebsara}}\ and\ \bibinfo {author} {\bibfnamefont {D.}~\bibnamefont
  {S\'en\'echal}},\ }\href {https://doi.org/10.1103/PhysRevLett.100.136402}
  {\bibfield  {journal} {\bibinfo  {journal} {Phys. Rev. Lett.}\ }\textbf
  {\bibinfo {volume} {100}},\ \bibinfo {pages} {136402} (\bibinfo {year}
  {2008})}\BibitemShut {NoStop}%
\bibitem [{\citenamefont {Yoshioka}\ \emph {et~al.}(2009)\citenamefont
  {Yoshioka}, \citenamefont {Koga},\ and\ \citenamefont
  {Kawakami}}]{Yoshioka2009}%
  \BibitemOpen
  \bibfield  {author} {\bibinfo {author} {\bibfnamefont {T.}~\bibnamefont
  {Yoshioka}}, \bibinfo {author} {\bibfnamefont {A.}~\bibnamefont {Koga}},\
  and\ \bibinfo {author} {\bibfnamefont {N.}~\bibnamefont {Kawakami}},\ }\href
  {https://doi.org/10.1103/PhysRevLett.103.036401} {\bibfield  {journal}
  {\bibinfo  {journal} {Phys. Rev. Lett.}\ }\textbf {\bibinfo {volume} {103}},\
  \bibinfo {pages} {036401} (\bibinfo {year} {2009})}\BibitemShut {NoStop}%
\bibitem [{\citenamefont {Yang}\ \emph {et~al.}(2010)\citenamefont {Yang},
  \citenamefont {L\"auchli}, \citenamefont {Mila},\ and\ \citenamefont
  {Schmidt}}]{Yang2010}%
  \BibitemOpen
  \bibfield  {author} {\bibinfo {author} {\bibfnamefont {H.-Y.}\ \bibnamefont
  {Yang}}, \bibinfo {author} {\bibfnamefont {A.~M.}\ \bibnamefont {L\"auchli}},
  \bibinfo {author} {\bibfnamefont {F.}~\bibnamefont {Mila}},\ and\ \bibinfo
  {author} {\bibfnamefont {K.~P.}\ \bibnamefont {Schmidt}},\ }\href
  {https://doi.org/10.1103/PhysRevLett.105.267204} {\bibfield  {journal}
  {\bibinfo  {journal} {Phys. Rev. Lett.}\ }\textbf {\bibinfo {volume} {105}},\
  \bibinfo {pages} {267204} (\bibinfo {year} {2010})}\BibitemShut {NoStop}%
\bibitem [{\citenamefont {Shirakawa}\ \emph {et~al.}(2017)\citenamefont
  {Shirakawa}, \citenamefont {Tohyama}, \citenamefont {Kokalj}, \citenamefont
  {Sota},\ and\ \citenamefont {Yunoki}}]{Shirakawa2017}%
  \BibitemOpen
  \bibfield  {author} {\bibinfo {author} {\bibfnamefont {T.}~\bibnamefont
  {Shirakawa}}, \bibinfo {author} {\bibfnamefont {T.}~\bibnamefont {Tohyama}},
  \bibinfo {author} {\bibfnamefont {J.}~\bibnamefont {Kokalj}}, \bibinfo
  {author} {\bibfnamefont {S.}~\bibnamefont {Sota}},\ and\ \bibinfo {author}
  {\bibfnamefont {S.}~\bibnamefont {Yunoki}},\ }\href
  {https://doi.org/10.1103/PhysRevB.96.205130} {\bibfield  {journal} {\bibinfo
  {journal} {Phys. Rev. B}\ }\textbf {\bibinfo {volume} {96}},\ \bibinfo
  {pages} {205130} (\bibinfo {year} {2017})}\BibitemShut {NoStop}%
\bibitem [{\citenamefont {Szasz}\ \emph {et~al.}(2020)\citenamefont {Szasz},
  \citenamefont {Motruk}, \citenamefont {Zaletel},\ and\ \citenamefont
  {Moore}}]{Szasz2020}%
  \BibitemOpen
  \bibfield  {author} {\bibinfo {author} {\bibfnamefont {A.}~\bibnamefont
  {Szasz}}, \bibinfo {author} {\bibfnamefont {J.}~\bibnamefont {Motruk}},
  \bibinfo {author} {\bibfnamefont {M.~P.}\ \bibnamefont {Zaletel}},\ and\
  \bibinfo {author} {\bibfnamefont {J.~E.}\ \bibnamefont {Moore}},\ }\href
  {https://doi.org/10.1103/PhysRevX.10.021042} {\bibfield  {journal} {\bibinfo
  {journal} {Phys. Rev. X}\ }\textbf {\bibinfo {volume} {10}},\ \bibinfo
  {pages} {021042} (\bibinfo {year} {2020})}\BibitemShut {NoStop}%
\bibitem [{\citenamefont {Wietek}\ \emph {et~al.}(2021)\citenamefont {Wietek},
  \citenamefont {Rossi}, \citenamefont {\ifmmode~\check{S}\else
  \v{S}\fi{}imkovic}, \citenamefont {Klett}, \citenamefont {Hansmann},
  \citenamefont {Ferrero}, \citenamefont {Stoudenmire}, \citenamefont
  {Sch\"afer},\ and\ \citenamefont {Georges}}]{Wietek2021}%
  \BibitemOpen
  \bibfield  {author} {\bibinfo {author} {\bibfnamefont {A.}~\bibnamefont
  {Wietek}}, \bibinfo {author} {\bibfnamefont {R.}~\bibnamefont {Rossi}},
  \bibinfo {author} {\bibfnamefont {F.}~\bibnamefont {\ifmmode~\check{S}\else
  \v{S}\fi{}imkovic}}, \bibinfo {author} {\bibfnamefont {M.}~\bibnamefont
  {Klett}}, \bibinfo {author} {\bibfnamefont {P.}~\bibnamefont {Hansmann}},
  \bibinfo {author} {\bibfnamefont {M.}~\bibnamefont {Ferrero}}, \bibinfo
  {author} {\bibfnamefont {E.~M.}\ \bibnamefont {Stoudenmire}}, \bibinfo
  {author} {\bibfnamefont {T.}~\bibnamefont {Sch\"afer}},\ and\ \bibinfo
  {author} {\bibfnamefont {A.}~\bibnamefont {Georges}},\ }\href
  {https://doi.org/10.1103/PhysRevX.11.041013} {\bibfield  {journal} {\bibinfo
  {journal} {Phys. Rev. X}\ }\textbf {\bibinfo {volume} {11}},\ \bibinfo
  {pages} {041013} (\bibinfo {year} {2021})}\BibitemShut {NoStop}%
\bibitem [{\citenamefont {Szasz}\ and\ \citenamefont
  {Motruk}(2021)}]{PhysRevB.103.235132}%
  \BibitemOpen
  \bibfield  {author} {\bibinfo {author} {\bibfnamefont {A.}~\bibnamefont
  {Szasz}}\ and\ \bibinfo {author} {\bibfnamefont {J.}~\bibnamefont {Motruk}},\
  }\href {https://doi.org/10.1103/PhysRevB.103.235132} {\bibfield  {journal}
  {\bibinfo  {journal} {Phys. Rev. B}\ }\textbf {\bibinfo {volume} {103}},\
  \bibinfo {pages} {235132} (\bibinfo {year} {2021})}\BibitemShut {NoStop}%
\bibitem [{\citenamefont {Zhu}\ \emph {et~al.}(2022)\citenamefont {Zhu},
  \citenamefont {Sheng},\ and\ \citenamefont {Vishwanath}}]{zhu2022doped}%
  \BibitemOpen
  \bibfield  {author} {\bibinfo {author} {\bibfnamefont {Z.}~\bibnamefont
  {Zhu}}, \bibinfo {author} {\bibfnamefont {D.~N.}\ \bibnamefont {Sheng}},\
  and\ \bibinfo {author} {\bibfnamefont {A.}~\bibnamefont {Vishwanath}},\
  }\href {https://doi.org/10.1103/PhysRevB.105.205110} {\bibfield  {journal}
  {\bibinfo  {journal} {Phys. Rev. B}\ }\textbf {\bibinfo {volume} {105}},\
  \bibinfo {pages} {205110} (\bibinfo {year} {2022})}\BibitemShut {NoStop}%
\bibitem [{\citenamefont {Chen}\ \emph {et~al.}(2022)\citenamefont {Chen},
  \citenamefont {Chen}, \citenamefont {Gong}, \citenamefont {Sheng},
  \citenamefont {Li},\ and\ \citenamefont {Weichselbaum}}]{chen2022quantum}%
  \BibitemOpen
  \bibfield  {author} {\bibinfo {author} {\bibfnamefont {B.-B.}\ \bibnamefont
  {Chen}}, \bibinfo {author} {\bibfnamefont {Z.}~\bibnamefont {Chen}}, \bibinfo
  {author} {\bibfnamefont {S.-S.}\ \bibnamefont {Gong}}, \bibinfo {author}
  {\bibfnamefont {D.~N.}\ \bibnamefont {Sheng}}, \bibinfo {author}
  {\bibfnamefont {W.}~\bibnamefont {Li}},\ and\ \bibinfo {author}
  {\bibfnamefont {A.}~\bibnamefont {Weichselbaum}},\ }\href
  {https://doi.org/10.1103/PhysRevB.106.094420} {\bibfield  {journal} {\bibinfo
   {journal} {Phys. Rev. B}\ }\textbf {\bibinfo {volume} {106}},\ \bibinfo
  {pages} {094420} (\bibinfo {year} {2022})}\BibitemShut {NoStop}%
\bibitem [{\citenamefont {Cookmeyer}\ \emph {et~al.}(2021)\citenamefont
  {Cookmeyer}, \citenamefont {Motruk},\ and\ \citenamefont
  {Moore}}]{PhysRevLett.127.087201}%
  \BibitemOpen
  \bibfield  {author} {\bibinfo {author} {\bibfnamefont {T.}~\bibnamefont
  {Cookmeyer}}, \bibinfo {author} {\bibfnamefont {J.}~\bibnamefont {Motruk}},\
  and\ \bibinfo {author} {\bibfnamefont {J.~E.}\ \bibnamefont {Moore}},\ }\href
  {https://doi.org/10.1103/PhysRevLett.127.087201} {\bibfield  {journal}
  {\bibinfo  {journal} {Phys. Rev. Lett.}\ }\textbf {\bibinfo {volume} {127}},\
  \bibinfo {pages} {087201} (\bibinfo {year} {2021})}\BibitemShut {NoStop}%
\bibitem [{\citenamefont {Jolicoeur}\ \emph {et~al.}(1990)\citenamefont
  {Jolicoeur}, \citenamefont {Dagotto}, \citenamefont {Gagliano},\ and\
  \citenamefont {Bacci}}]{PhysRevB.42.4800}%
  \BibitemOpen
  \bibfield  {author} {\bibinfo {author} {\bibfnamefont {T.}~\bibnamefont
  {Jolicoeur}}, \bibinfo {author} {\bibfnamefont {E.}~\bibnamefont {Dagotto}},
  \bibinfo {author} {\bibfnamefont {E.}~\bibnamefont {Gagliano}},\ and\
  \bibinfo {author} {\bibfnamefont {S.}~\bibnamefont {Bacci}},\ }\href
  {https://doi.org/10.1103/PhysRevB.42.4800} {\bibfield  {journal} {\bibinfo
  {journal} {Phys. Rev. B}\ }\textbf {\bibinfo {volume} {42}},\ \bibinfo
  {pages} {4800} (\bibinfo {year} {1990})}\BibitemShut {NoStop}%
\bibitem [{\citenamefont {Villain}\ \emph {et~al.}(1980)\citenamefont
  {Villain}, \citenamefont {Bidaux}, \citenamefont {Carton},\ and\
  \citenamefont {Conte}}]{villain1980order}%
  \BibitemOpen
  \bibfield  {author} {\bibinfo {author} {\bibfnamefont {J.}~\bibnamefont
  {Villain}}, \bibinfo {author} {\bibfnamefont {R.}~\bibnamefont {Bidaux}},
  \bibinfo {author} {\bibfnamefont {J.-P.}\ \bibnamefont {Carton}},\ and\
  \bibinfo {author} {\bibfnamefont {R.}~\bibnamefont {Conte}},\ }\href@noop {}
  {\bibfield  {journal} {\bibinfo  {journal} {Journal de Physique}\ }\textbf
  {\bibinfo {volume} {41}},\ \bibinfo {pages} {1263} (\bibinfo {year}
  {1980})}\BibitemShut {NoStop}%
\bibitem [{\citenamefont {Henley}(1989)}]{henley1989ordering}%
  \BibitemOpen
  \bibfield  {author} {\bibinfo {author} {\bibfnamefont {C.~L.}\ \bibnamefont
  {Henley}},\ }\href@noop {} {\bibfield  {journal} {\bibinfo  {journal}
  {Physical review letters}\ }\textbf {\bibinfo {volume} {62}},\ \bibinfo
  {pages} {2056} (\bibinfo {year} {1989})}\BibitemShut {NoStop}%
\bibitem [{\citenamefont {Chubukov}\ and\ \citenamefont
  {Jolicoeur}(1992)}]{Chubukov1992}%
  \BibitemOpen
  \bibfield  {author} {\bibinfo {author} {\bibfnamefont {A.~V.}\ \bibnamefont
  {Chubukov}}\ and\ \bibinfo {author} {\bibfnamefont {T.}~\bibnamefont
  {Jolicoeur}},\ }\href {https://doi.org/10.1103/PhysRevB.46.11137} {\bibfield
  {journal} {\bibinfo  {journal} {Phys. Rev. B}\ }\textbf {\bibinfo {volume}
  {46}},\ \bibinfo {pages} {11137} (\bibinfo {year} {1992})}\BibitemShut
  {NoStop}%
\bibitem [{\citenamefont {Kaneko}\ \emph {et~al.}(2014)\citenamefont {Kaneko},
  \citenamefont {Morita},\ and\ \citenamefont {Imada}}]{Kaneko2014}%
  \BibitemOpen
  \bibfield  {author} {\bibinfo {author} {\bibfnamefont {R.}~\bibnamefont
  {Kaneko}}, \bibinfo {author} {\bibfnamefont {S.}~\bibnamefont {Morita}},\
  and\ \bibinfo {author} {\bibfnamefont {M.}~\bibnamefont {Imada}},\ }\href
  {https://doi.org/10.7566/JPSJ.83.093707} {\bibfield  {journal} {\bibinfo
  {journal} {Journal of the Physical Society of Japan}\ }\textbf {\bibinfo
  {volume} {83}},\ \bibinfo {pages} {093707} (\bibinfo {year}
  {2014})}\BibitemShut {NoStop}%
\bibitem [{\citenamefont {Iqbal}\ \emph {et~al.}(2016)\citenamefont {Iqbal},
  \citenamefont {Hu}, \citenamefont {Thomale}, \citenamefont {Poilblanc},\ and\
  \citenamefont {Becca}}]{Iqbal2016}%
  \BibitemOpen
  \bibfield  {author} {\bibinfo {author} {\bibfnamefont {Y.}~\bibnamefont
  {Iqbal}}, \bibinfo {author} {\bibfnamefont {W.-J.}\ \bibnamefont {Hu}},
  \bibinfo {author} {\bibfnamefont {R.}~\bibnamefont {Thomale}}, \bibinfo
  {author} {\bibfnamefont {D.}~\bibnamefont {Poilblanc}},\ and\ \bibinfo
  {author} {\bibfnamefont {F.}~\bibnamefont {Becca}},\ }\href
  {https://doi.org/10.1103/PhysRevB.93.144411} {\bibfield  {journal} {\bibinfo
  {journal} {Phys. Rev. B}\ }\textbf {\bibinfo {volume} {93}},\ \bibinfo
  {pages} {144411} (\bibinfo {year} {2016})}\BibitemShut {NoStop}%
\bibitem [{\citenamefont {Zhu}\ and\ \citenamefont {White}(2015)}]{Zhu2015}%
  \BibitemOpen
  \bibfield  {author} {\bibinfo {author} {\bibfnamefont {Z.}~\bibnamefont
  {Zhu}}\ and\ \bibinfo {author} {\bibfnamefont {S.~R.}\ \bibnamefont
  {White}},\ }\href {https://doi.org/10.1103/PhysRevB.92.041105} {\bibfield
  {journal} {\bibinfo  {journal} {Phys. Rev. B}\ }\textbf {\bibinfo {volume}
  {92}},\ \bibinfo {pages} {041105} (\bibinfo {year} {2015})}\BibitemShut
  {NoStop}%
\bibitem [{\citenamefont {Hu}\ \emph {et~al.}(2015)\citenamefont {Hu},
  \citenamefont {Gong}, \citenamefont {Zhu},\ and\ \citenamefont
  {Sheng}}]{Hu2015}%
  \BibitemOpen
  \bibfield  {author} {\bibinfo {author} {\bibfnamefont {W.-J.}\ \bibnamefont
  {Hu}}, \bibinfo {author} {\bibfnamefont {S.-S.}\ \bibnamefont {Gong}},
  \bibinfo {author} {\bibfnamefont {W.}~\bibnamefont {Zhu}},\ and\ \bibinfo
  {author} {\bibfnamefont {D.~N.}\ \bibnamefont {Sheng}},\ }\href
  {https://doi.org/10.1103/PhysRevB.92.140403} {\bibfield  {journal} {\bibinfo
  {journal} {Phys. Rev. B}\ }\textbf {\bibinfo {volume} {92}},\ \bibinfo
  {pages} {140403} (\bibinfo {year} {2015})}\BibitemShut {NoStop}%
\bibitem [{\citenamefont {Gong}\ \emph
  {et~al.}(2017{\natexlab{a}})\citenamefont {Gong}, \citenamefont {Zhu},
  \citenamefont {Zhu}, \citenamefont {Sheng},\ and\ \citenamefont
  {Yang}}]{Gong2017}%
  \BibitemOpen
  \bibfield  {author} {\bibinfo {author} {\bibfnamefont {S.-S.}\ \bibnamefont
  {Gong}}, \bibinfo {author} {\bibfnamefont {W.}~\bibnamefont {Zhu}}, \bibinfo
  {author} {\bibfnamefont {J.-X.}\ \bibnamefont {Zhu}}, \bibinfo {author}
  {\bibfnamefont {D.~N.}\ \bibnamefont {Sheng}},\ and\ \bibinfo {author}
  {\bibfnamefont {K.}~\bibnamefont {Yang}},\ }\href
  {https://doi.org/10.1103/PhysRevB.96.075116} {\bibfield  {journal} {\bibinfo
  {journal} {Phys. Rev. B}\ }\textbf {\bibinfo {volume} {96}},\ \bibinfo
  {pages} {075116} (\bibinfo {year} {2017}{\natexlab{a}})}\BibitemShut
  {NoStop}%
\bibitem [{\citenamefont {Hu}\ \emph {et~al.}(2019)\citenamefont {Hu},
  \citenamefont {Zhu}, \citenamefont {Eggert},\ and\ \citenamefont
  {He}}]{Hu2019}%
  \BibitemOpen
  \bibfield  {author} {\bibinfo {author} {\bibfnamefont {S.}~\bibnamefont
  {Hu}}, \bibinfo {author} {\bibfnamefont {W.}~\bibnamefont {Zhu}}, \bibinfo
  {author} {\bibfnamefont {S.}~\bibnamefont {Eggert}},\ and\ \bibinfo {author}
  {\bibfnamefont {Y.-C.}\ \bibnamefont {He}},\ }\href
  {https://doi.org/10.1103/PhysRevLett.123.207203} {\bibfield  {journal}
  {\bibinfo  {journal} {Phys. Rev. Lett.}\ }\textbf {\bibinfo {volume} {123}},\
  \bibinfo {pages} {207203} (\bibinfo {year} {2019})}\BibitemShut {NoStop}%
\bibitem [{\citenamefont {Drescher}\ \emph {et~al.}(2022)\citenamefont
  {Drescher}, \citenamefont {Vanderstraeten}, \citenamefont {Moessner},\ and\
  \citenamefont {Pollmann}}]{drescher2022dynamical}%
  \BibitemOpen
  \bibfield  {author} {\bibinfo {author} {\bibfnamefont {M.}~\bibnamefont
  {Drescher}}, \bibinfo {author} {\bibfnamefont {L.}~\bibnamefont
  {Vanderstraeten}}, \bibinfo {author} {\bibfnamefont {R.}~\bibnamefont
  {Moessner}},\ and\ \bibinfo {author} {\bibfnamefont {F.}~\bibnamefont
  {Pollmann}},\ }\href@noop {} {\bibinfo {title} {Dynamical signatures of
  symmetry broken and liquid phases in an $s=1/2$ heisenberg antiferromagnet on
  the triangular lattice}} (\bibinfo {year} {2022}),\ \Eprint
  {https://arxiv.org/abs/2209.03344} {arXiv:2209.03344 [cond-mat.str-el]}
  \BibitemShut {NoStop}%
\bibitem [{\citenamefont {Misumi}\ \emph {et~al.}(2017)\citenamefont {Misumi},
  \citenamefont {Kaneko},\ and\ \citenamefont {Ohta}}]{Misumi2017}%
  \BibitemOpen
  \bibfield  {author} {\bibinfo {author} {\bibfnamefont {K.}~\bibnamefont
  {Misumi}}, \bibinfo {author} {\bibfnamefont {T.}~\bibnamefont {Kaneko}},\
  and\ \bibinfo {author} {\bibfnamefont {Y.}~\bibnamefont {Ohta}},\ }\href
  {https://doi.org/10.1103/PhysRevB.95.075124} {\bibfield  {journal} {\bibinfo
  {journal} {Phys. Rev. B}\ }\textbf {\bibinfo {volume} {95}},\ \bibinfo
  {pages} {075124} (\bibinfo {year} {2017})}\BibitemShut {NoStop}%
\bibitem [{\citenamefont {Tocchio}\ \emph {et~al.}(2020)\citenamefont
  {Tocchio}, \citenamefont {Montorsi},\ and\ \citenamefont
  {Becca}}]{Tocchio2020}%
  \BibitemOpen
  \bibfield  {author} {\bibinfo {author} {\bibfnamefont {L.~F.}\ \bibnamefont
  {Tocchio}}, \bibinfo {author} {\bibfnamefont {A.}~\bibnamefont {Montorsi}},\
  and\ \bibinfo {author} {\bibfnamefont {F.}~\bibnamefont {Becca}},\ }\href
  {https://doi.org/10.1103/PhysRevB.102.115150} {\bibfield  {journal} {\bibinfo
   {journal} {Phys. Rev. B}\ }\textbf {\bibinfo {volume} {102}},\ \bibinfo
  {pages} {115150} (\bibinfo {year} {2020})}\BibitemShut {NoStop}%
\bibitem [{\citenamefont {Morita}\ \emph {et~al.}(2002)\citenamefont {Morita},
  \citenamefont {Watanabe},\ and\ \citenamefont
  {Imada}}]{doi:10.1143/JPSJ.71.2109}%
  \BibitemOpen
  \bibfield  {author} {\bibinfo {author} {\bibfnamefont {H.}~\bibnamefont
  {Morita}}, \bibinfo {author} {\bibfnamefont {S.}~\bibnamefont {Watanabe}},\
  and\ \bibinfo {author} {\bibfnamefont {M.}~\bibnamefont {Imada}},\ }\href
  {https://doi.org/10.1143/JPSJ.71.2109} {\bibfield  {journal} {\bibinfo
  {journal} {Journal of the Physical Society of Japan}\ }\textbf {\bibinfo
  {volume} {71}},\ \bibinfo {pages} {2109} (\bibinfo {year}
  {2002})}\BibitemShut {NoStop}%
\bibitem [{\citenamefont {Watanabe}\ \emph {et~al.}(2008)\citenamefont
  {Watanabe}, \citenamefont {Yokoyama}, \citenamefont {Tanaka},\ and\
  \citenamefont {Inoue}}]{PhysRevB.77.214505}%
  \BibitemOpen
  \bibfield  {author} {\bibinfo {author} {\bibfnamefont {T.}~\bibnamefont
  {Watanabe}}, \bibinfo {author} {\bibfnamefont {H.}~\bibnamefont {Yokoyama}},
  \bibinfo {author} {\bibfnamefont {Y.}~\bibnamefont {Tanaka}},\ and\ \bibinfo
  {author} {\bibfnamefont {J.}~\bibnamefont {Inoue}},\ }\href
  {https://doi.org/10.1103/PhysRevB.77.214505} {\bibfield  {journal} {\bibinfo
  {journal} {Phys. Rev. B}\ }\textbf {\bibinfo {volume} {77}},\ \bibinfo
  {pages} {214505} (\bibinfo {year} {2008})}\BibitemShut {NoStop}%
\bibitem [{\citenamefont {Tocchio}\ \emph {et~al.}(2009)\citenamefont
  {Tocchio}, \citenamefont {Parola}, \citenamefont {Gros},\ and\ \citenamefont
  {Becca}}]{PhysRevB.80.064419}%
  \BibitemOpen
  \bibfield  {author} {\bibinfo {author} {\bibfnamefont {L.~F.}\ \bibnamefont
  {Tocchio}}, \bibinfo {author} {\bibfnamefont {A.}~\bibnamefont {Parola}},
  \bibinfo {author} {\bibfnamefont {C.}~\bibnamefont {Gros}},\ and\ \bibinfo
  {author} {\bibfnamefont {F.}~\bibnamefont {Becca}},\ }\href
  {https://doi.org/10.1103/PhysRevB.80.064419} {\bibfield  {journal} {\bibinfo
  {journal} {Phys. Rev. B}\ }\textbf {\bibinfo {volume} {80}},\ \bibinfo
  {pages} {064419} (\bibinfo {year} {2009})}\BibitemShut {NoStop}%
\bibitem [{\citenamefont {Tocchio}\ \emph {et~al.}(2013)\citenamefont
  {Tocchio}, \citenamefont {Feldner}, \citenamefont {Becca}, \citenamefont
  {Valent\'{\i}},\ and\ \citenamefont {Gros}}]{PhysRevB.87.035143}%
  \BibitemOpen
  \bibfield  {author} {\bibinfo {author} {\bibfnamefont {L.~F.}\ \bibnamefont
  {Tocchio}}, \bibinfo {author} {\bibfnamefont {H.}~\bibnamefont {Feldner}},
  \bibinfo {author} {\bibfnamefont {F.}~\bibnamefont {Becca}}, \bibinfo
  {author} {\bibfnamefont {R.}~\bibnamefont {Valent\'{\i}}},\ and\ \bibinfo
  {author} {\bibfnamefont {C.}~\bibnamefont {Gros}},\ }\href
  {https://doi.org/10.1103/PhysRevB.87.035143} {\bibfield  {journal} {\bibinfo
  {journal} {Phys. Rev. B}\ }\textbf {\bibinfo {volume} {87}},\ \bibinfo
  {pages} {035143} (\bibinfo {year} {2013})}\BibitemShut {NoStop}%
\bibitem [{\citenamefont {Yu}\ \emph {et~al.}(2022)\citenamefont {Yu},
  \citenamefont {Li}, \citenamefont {Iskakov},\ and\ \citenamefont
  {Gull}}]{arxiv.2211.09234}%
  \BibitemOpen
  \bibfield  {author} {\bibinfo {author} {\bibfnamefont {Y.}~\bibnamefont
  {Yu}}, \bibinfo {author} {\bibfnamefont {S.}~\bibnamefont {Li}}, \bibinfo
  {author} {\bibfnamefont {S.}~\bibnamefont {Iskakov}},\ and\ \bibinfo {author}
  {\bibfnamefont {E.}~\bibnamefont {Gull}},\ }\href
  {https://doi.org/10.48550/ARXIV.2211.09234} {\bibinfo {title} {Magnetic
  phases of the anisotropic triangular lattice hubbard model}} (\bibinfo {year}
  {2022})\BibitemShut {NoStop}%
\bibitem [{\citenamefont {Sherman}\ \emph {et~al.}(2022)\citenamefont
  {Sherman}, \citenamefont {Dupont},\ and\ \citenamefont
  {Moore}}]{sherman2022spectral}%
  \BibitemOpen
  \bibfield  {author} {\bibinfo {author} {\bibfnamefont {N.~E.}\ \bibnamefont
  {Sherman}}, \bibinfo {author} {\bibfnamefont {M.}~\bibnamefont {Dupont}},\
  and\ \bibinfo {author} {\bibfnamefont {J.~E.}\ \bibnamefont {Moore}},\
  }\href@noop {} {\bibinfo {title} {Spectral function of the $j_1-j_2$
  heisenberg model on the triangular lattice}} (\bibinfo {year} {2022}),\
  \Eprint {https://arxiv.org/abs/2209.00739} {arXiv:2209.00739
  [cond-mat.str-el]} \BibitemShut {NoStop}%
\bibitem [{\citenamefont {Kadow}\ \emph {et~al.}(2022)\citenamefont {Kadow},
  \citenamefont {Vanderstraeten},\ and\ \citenamefont
  {Knap}}]{PhysRevB.106.094417}%
  \BibitemOpen
  \bibfield  {author} {\bibinfo {author} {\bibfnamefont {W.}~\bibnamefont
  {Kadow}}, \bibinfo {author} {\bibfnamefont {L.}~\bibnamefont
  {Vanderstraeten}},\ and\ \bibinfo {author} {\bibfnamefont {M.}~\bibnamefont
  {Knap}},\ }\href {https://doi.org/10.1103/PhysRevB.106.094417} {\bibfield
  {journal} {\bibinfo  {journal} {Phys. Rev. B}\ }\textbf {\bibinfo {volume}
  {106}},\ \bibinfo {pages} {094417} (\bibinfo {year} {2022})}\BibitemShut
  {NoStop}%
\bibitem [{\citenamefont {Schrieffer}\ \emph {et~al.}(1989)\citenamefont
  {Schrieffer}, \citenamefont {Wen},\ and\ \citenamefont
  {Zhang}}]{PhysRevB.39.11663}%
  \BibitemOpen
  \bibfield  {author} {\bibinfo {author} {\bibfnamefont {J.~R.}\ \bibnamefont
  {Schrieffer}}, \bibinfo {author} {\bibfnamefont {X.~G.}\ \bibnamefont
  {Wen}},\ and\ \bibinfo {author} {\bibfnamefont {S.~C.}\ \bibnamefont
  {Zhang}},\ }\href {https://doi.org/10.1103/PhysRevB.39.11663} {\bibfield
  {journal} {\bibinfo  {journal} {Phys. Rev. B}\ }\textbf {\bibinfo {volume}
  {39}},\ \bibinfo {pages} {11663} (\bibinfo {year} {1989})}\BibitemShut
  {NoStop}%
\bibitem [{\citenamefont {Chubukov}\ and\ \citenamefont
  {Frenkel}(1992)}]{PhysRevB.46.11884}%
  \BibitemOpen
  \bibfield  {author} {\bibinfo {author} {\bibfnamefont {A.~V.}\ \bibnamefont
  {Chubukov}}\ and\ \bibinfo {author} {\bibfnamefont {D.~M.}\ \bibnamefont
  {Frenkel}},\ }\href {https://doi.org/10.1103/PhysRevB.46.11884} {\bibfield
  {journal} {\bibinfo  {journal} {Phys. Rev. B}\ }\textbf {\bibinfo {volume}
  {46}},\ \bibinfo {pages} {11884} (\bibinfo {year} {1992})}\BibitemShut
  {NoStop}%
\bibitem [{\citenamefont {Singh}\ and\ \citenamefont {Te\ifmmode \check{s}\else
  \v{s}\fi{}anovi\ifmmode~\acute{c}\else \'{c}\fi{}}(1990)}]{PhysRevB.41.614}%
  \BibitemOpen
  \bibfield  {author} {\bibinfo {author} {\bibfnamefont {A.}~\bibnamefont
  {Singh}}\ and\ \bibinfo {author} {\bibfnamefont {Z.}~\bibnamefont {Te\ifmmode
  \check{s}\else \v{s}\fi{}anovi\ifmmode~\acute{c}\else \'{c}\fi{}}},\ }\href
  {https://doi.org/10.1103/PhysRevB.41.614} {\bibfield  {journal} {\bibinfo
  {journal} {Phys. Rev. B}\ }\textbf {\bibinfo {volume} {41}},\ \bibinfo
  {pages} {614} (\bibinfo {year} {1990})}\BibitemShut {NoStop}%
\bibitem [{\citenamefont {Peres}\ and\ \citenamefont
  {Ara\'ujo}(2002)}]{peres2002spin}%
  \BibitemOpen
  \bibfield  {author} {\bibinfo {author} {\bibfnamefont {N.~M.~R.}\
  \bibnamefont {Peres}}\ and\ \bibinfo {author} {\bibfnamefont {M.~A.~N.}\
  \bibnamefont {Ara\'ujo}},\ }\href
  {https://doi.org/10.1103/PhysRevB.65.132404} {\bibfield  {journal} {\bibinfo
  {journal} {Phys. Rev. B}\ }\textbf {\bibinfo {volume} {65}},\ \bibinfo
  {pages} {132404} (\bibinfo {year} {2002})}\BibitemShut {NoStop}%
\bibitem [{\citenamefont {Peres}\ and\ \citenamefont
  {Araújo}(2003)}]{peres2003spin}%
  \BibitemOpen
  \bibfield  {author} {\bibinfo {author} {\bibfnamefont {N.~M.~R.}\
  \bibnamefont {Peres}}\ and\ \bibinfo {author} {\bibfnamefont {M.~A.~N.}\
  \bibnamefont {Araújo}},\ }\href
  {https://doi.org/https://doi.org/10.1002/pssb.200301719} {\bibfield
  {journal} {\bibinfo  {journal} {physica status solidi (b)}\ }\textbf
  {\bibinfo {volume} {236}},\ \bibinfo {pages} {523} (\bibinfo {year}
  {2003})},\ \Eprint
  {https://arxiv.org/abs/https://onlinelibrary.wiley.com/doi/pdf/10.1002/pssb.200301719}
  {https://onlinelibrary.wiley.com/doi/pdf/10.1002/pssb.200301719} \BibitemShut
  {NoStop}%
\bibitem [{\citenamefont {Knolle}\ \emph {et~al.}(2010)\citenamefont {Knolle},
  \citenamefont {Eremin}, \citenamefont {Chubukov},\ and\ \citenamefont
  {Moessner}}]{knolle2010theory}%
  \BibitemOpen
  \bibfield  {author} {\bibinfo {author} {\bibfnamefont {J.}~\bibnamefont
  {Knolle}}, \bibinfo {author} {\bibfnamefont {I.}~\bibnamefont {Eremin}},
  \bibinfo {author} {\bibfnamefont {A.~V.}\ \bibnamefont {Chubukov}},\ and\
  \bibinfo {author} {\bibfnamefont {R.}~\bibnamefont {Moessner}},\ }\href
  {https://doi.org/10.1103/PhysRevB.81.140506} {\bibfield  {journal} {\bibinfo
  {journal} {Phys. Rev. B}\ }\textbf {\bibinfo {volume} {81}},\ \bibinfo
  {pages} {140506} (\bibinfo {year} {2010})}\BibitemShut {NoStop}%
\bibitem [{\citenamefont {Brydon}\ and\ \citenamefont
  {Timm}(2009)}]{brydon2009spin}%
  \BibitemOpen
  \bibfield  {author} {\bibinfo {author} {\bibfnamefont {P.~M.~R.}\
  \bibnamefont {Brydon}}\ and\ \bibinfo {author} {\bibfnamefont
  {C.}~\bibnamefont {Timm}},\ }\href
  {https://doi.org/10.1103/PhysRevB.80.174401} {\bibfield  {journal} {\bibinfo
  {journal} {Phys. Rev. B}\ }\textbf {\bibinfo {volume} {80}},\ \bibinfo
  {pages} {174401} (\bibinfo {year} {2009})}\BibitemShut {NoStop}%
\bibitem [{\citenamefont {Kaneshita}\ and\ \citenamefont
  {Tohyama}(2010)}]{kaneshita2010spin}%
  \BibitemOpen
  \bibfield  {author} {\bibinfo {author} {\bibfnamefont {E.}~\bibnamefont
  {Kaneshita}}\ and\ \bibinfo {author} {\bibfnamefont {T.}~\bibnamefont
  {Tohyama}},\ }\href {https://doi.org/10.1103/PhysRevB.82.094441} {\bibfield
  {journal} {\bibinfo  {journal} {Phys. Rev. B}\ }\textbf {\bibinfo {volume}
  {82}},\ \bibinfo {pages} {094441} (\bibinfo {year} {2010})}\BibitemShut
  {NoStop}%
\bibitem [{\citenamefont {Knolle}\ \emph {et~al.}(2011)\citenamefont {Knolle},
  \citenamefont {Eremin},\ and\ \citenamefont
  {Moessner}}]{knolle2011multiorbital}%
  \BibitemOpen
  \bibfield  {author} {\bibinfo {author} {\bibfnamefont {J.}~\bibnamefont
  {Knolle}}, \bibinfo {author} {\bibfnamefont {I.}~\bibnamefont {Eremin}},\
  and\ \bibinfo {author} {\bibfnamefont {R.}~\bibnamefont {Moessner}},\ }\href
  {https://doi.org/10.1103/PhysRevB.83.224503} {\bibfield  {journal} {\bibinfo
  {journal} {Phys. Rev. B}\ }\textbf {\bibinfo {volume} {83}},\ \bibinfo
  {pages} {224503} (\bibinfo {year} {2011})}\BibitemShut {NoStop}%
\bibitem [{\citenamefont {Lovesey}(1984)}]{lovesey1984theory}%
  \BibitemOpen
  \bibfield  {author} {\bibinfo {author} {\bibfnamefont {S.}~\bibnamefont
  {Lovesey}},\ }\href {https://books.google.de/books?id=MoK7wAEACAAJ} {\emph
  {\bibinfo {title} {Theory of Neutron Scattering from Condensed Matter}}},\
  \bibinfo {series} {International series of monographs on physics}\ No.\
  \bibinfo {number} {Bd. 2}\ (\bibinfo  {publisher} {Clarendon Press},\
  \bibinfo {year} {1984})\BibitemShut {NoStop}%
\bibitem [{\citenamefont {Moriya}(2012)}]{moriya2012spin}%
  \BibitemOpen
  \bibfield  {author} {\bibinfo {author} {\bibfnamefont {T.}~\bibnamefont
  {Moriya}},\ }\href@noop {} {\emph {\bibinfo {title} {Spin fluctuations in
  itinerant electron magnetism}}},\ Vol.~\bibinfo {volume} {56}\ (\bibinfo
  {publisher} {Springer Science \& Business Media},\ \bibinfo {year}
  {2012})\BibitemShut {NoStop}%
\bibitem [{\citenamefont {Singh}(1991)}]{PhysRevB.43.3617}%
  \BibitemOpen
  \bibfield  {author} {\bibinfo {author} {\bibfnamefont {A.}~\bibnamefont
  {Singh}},\ }\href {https://doi.org/10.1103/PhysRevB.43.3617} {\bibfield
  {journal} {\bibinfo  {journal} {Phys. Rev. B}\ }\textbf {\bibinfo {volume}
  {43}},\ \bibinfo {pages} {3617} (\bibinfo {year} {1991})}\BibitemShut
  {NoStop}%
\bibitem [{\citenamefont {Edwards}\ and\ \citenamefont
  {Hertz}(1973)}]{edwards1973electron}%
  \BibitemOpen
  \bibfield  {author} {\bibinfo {author} {\bibfnamefont {D.}~\bibnamefont
  {Edwards}}\ and\ \bibinfo {author} {\bibfnamefont {J.}~\bibnamefont
  {Hertz}},\ }\href@noop {} {\bibfield  {journal} {\bibinfo  {journal} {Journal
  of Physics F: Metal Physics}\ }\textbf {\bibinfo {volume} {3}},\ \bibinfo
  {pages} {2191} (\bibinfo {year} {1973})}\BibitemShut {NoStop}%
\bibitem [{\citenamefont {Ghosh}\ and\ \citenamefont
  {Singh}(2008)}]{PhysRevB.77.094430}%
  \BibitemOpen
  \bibfield  {author} {\bibinfo {author} {\bibfnamefont {S.}~\bibnamefont
  {Ghosh}}\ and\ \bibinfo {author} {\bibfnamefont {A.}~\bibnamefont {Singh}},\
  }\href {https://doi.org/10.1103/PhysRevB.77.094430} {\bibfield  {journal}
  {\bibinfo  {journal} {Phys. Rev. B}\ }\textbf {\bibinfo {volume} {77}},\
  \bibinfo {pages} {094430} (\bibinfo {year} {2008})}\BibitemShut {NoStop}%
\bibitem [{\citenamefont {Hickey}\ \emph {et~al.}(2017)\citenamefont {Hickey},
  \citenamefont {Cincio}, \citenamefont {Papi\ifmmode~\acute{c}\else
  \'{c}\fi{}},\ and\ \citenamefont {Paramekanti}}]{PhysRevB.96.115115}%
  \BibitemOpen
  \bibfield  {author} {\bibinfo {author} {\bibfnamefont {C.}~\bibnamefont
  {Hickey}}, \bibinfo {author} {\bibfnamefont {L.}~\bibnamefont {Cincio}},
  \bibinfo {author} {\bibfnamefont {Z.}~\bibnamefont
  {Papi\ifmmode~\acute{c}\else \'{c}\fi{}}},\ and\ \bibinfo {author}
  {\bibfnamefont {A.}~\bibnamefont {Paramekanti}},\ }\href
  {https://doi.org/10.1103/PhysRevB.96.115115} {\bibfield  {journal} {\bibinfo
  {journal} {Phys. Rev. B}\ }\textbf {\bibinfo {volume} {96}},\ \bibinfo
  {pages} {115115} (\bibinfo {year} {2017})}\BibitemShut {NoStop}%
\bibitem [{\citenamefont {Saadatmand}\ and\ \citenamefont
  {McCulloch}(2017)}]{saadatmand2017detection}%
  \BibitemOpen
  \bibfield  {author} {\bibinfo {author} {\bibfnamefont {S.~N.}\ \bibnamefont
  {Saadatmand}}\ and\ \bibinfo {author} {\bibfnamefont {I.~P.}\ \bibnamefont
  {McCulloch}},\ }\href {https://doi.org/10.1103/PhysRevB.96.075117} {\bibfield
   {journal} {\bibinfo  {journal} {Phys. Rev. B}\ }\textbf {\bibinfo {volume}
  {96}},\ \bibinfo {pages} {075117} (\bibinfo {year} {2017})}\BibitemShut
  {NoStop}%
\bibitem [{\citenamefont {Gong}\ \emph
  {et~al.}(2017{\natexlab{b}})\citenamefont {Gong}, \citenamefont {Zhu},
  \citenamefont {Zhu}, \citenamefont {Sheng},\ and\ \citenamefont
  {Yang}}]{gong2017global}%
  \BibitemOpen
  \bibfield  {author} {\bibinfo {author} {\bibfnamefont {S.-S.}\ \bibnamefont
  {Gong}}, \bibinfo {author} {\bibfnamefont {W.}~\bibnamefont {Zhu}}, \bibinfo
  {author} {\bibfnamefont {J.-X.}\ \bibnamefont {Zhu}}, \bibinfo {author}
  {\bibfnamefont {D.~N.}\ \bibnamefont {Sheng}},\ and\ \bibinfo {author}
  {\bibfnamefont {K.}~\bibnamefont {Yang}},\ }\href
  {https://doi.org/10.1103/PhysRevB.96.075116} {\bibfield  {journal} {\bibinfo
  {journal} {Phys. Rev. B}\ }\textbf {\bibinfo {volume} {96}},\ \bibinfo
  {pages} {075116} (\bibinfo {year} {2017}{\natexlab{b}})}\BibitemShut
  {NoStop}%
\bibitem [{\citenamefont {Ho}\ \emph {et~al.}(2001)\citenamefont {Ho},
  \citenamefont {Muthukumar}, \citenamefont {Ogata},\ and\ \citenamefont
  {Anderson}}]{ho2001nature}%
  \BibitemOpen
  \bibfield  {author} {\bibinfo {author} {\bibfnamefont {C.-M.}\ \bibnamefont
  {Ho}}, \bibinfo {author} {\bibfnamefont {V.~N.}\ \bibnamefont {Muthukumar}},
  \bibinfo {author} {\bibfnamefont {M.}~\bibnamefont {Ogata}},\ and\ \bibinfo
  {author} {\bibfnamefont {P.~W.}\ \bibnamefont {Anderson}},\ }\href
  {https://doi.org/10.1103/PhysRevLett.86.1626} {\bibfield  {journal} {\bibinfo
   {journal} {Phys. Rev. Lett.}\ }\textbf {\bibinfo {volume} {86}},\ \bibinfo
  {pages} {1626} (\bibinfo {year} {2001})}\BibitemShut {NoStop}%
\bibitem [{\citenamefont {Zhang}\ and\ \citenamefont
  {Li}(2020)}]{zhang2020resonating}%
  \BibitemOpen
  \bibfield  {author} {\bibinfo {author} {\bibfnamefont {C.}~\bibnamefont
  {Zhang}}\ and\ \bibinfo {author} {\bibfnamefont {T.}~\bibnamefont {Li}},\
  }\href {https://doi.org/10.1103/PhysRevB.102.075108} {\bibfield  {journal}
  {\bibinfo  {journal} {Phys. Rev. B}\ }\textbf {\bibinfo {volume} {102}},\
  \bibinfo {pages} {075108} (\bibinfo {year} {2020})}\BibitemShut {NoStop}%
\bibitem [{\citenamefont {Willsher}\ \emph {et~al.}(2022)\citenamefont
  {Willsher}, \citenamefont {Jin},\ and\ \citenamefont
  {Knolle}}]{willsher_josef_2022_7290111}%
  \BibitemOpen
  \bibfield  {author} {\bibinfo {author} {\bibfnamefont {J.}~\bibnamefont
  {Willsher}}, \bibinfo {author} {\bibfnamefont {H.-K.}\ \bibnamefont {Jin}},\
  and\ \bibinfo {author} {\bibfnamefont {J.}~\bibnamefont {Knolle}},\ }\href
  {https://doi.org/10.5281/zenodo.7290111} {\bibinfo {title} {{Magnetic
  excitations, phase diagram and order-by-disorder in the extended
  triangular-lattice Hubbard model}}} (\bibinfo {year} {2022})\BibitemShut
  {NoStop}%
\end{thebibliography}%


%

\onecolumngrid

\appendix

\section*{Appendix}
The stripe Hamiltonian is
\begin{equation}
    H_{\mathrm{stripe}}(\mathbf{k}) =
\begin{pmatrix}
\zeta_{\mathbf{k}}\mathbb{I}-\sigma\cdot \Delta_A & \delta_{\mathbf{k}}\mathbb{I} \\
\delta_{\mathbf{k}}^*\mathbb{I} & \zeta_{\mathbf{k}}\mathbb{I}-\sigma\cdot \Delta_B
\end{pmatrix}.
\end{equation}
In order to define and work with this matrix as a function of momentum, we take the inverse lattice as a basis of momentum $(k_1, k_2)$ such that if $\mathbf{r}_i$ are the lattice unit vectors, $\mathbf{k}\cdot \mathbf{r}_i = k_i$. Written as a function of $k_{1,2}$, the $AB$ hopping element is explicitly given by
\begin{equation}
    \delta_{\mathbf{k}} = t\left[1 + e^{-i k_1} + e^{-i k_2} + e^{-i (k_1+k_2)}\right]+ t' \left[e^{i k_1} + e^{i(k_1-k_2)} + e^{-i(2k_1+k_2)} + e^{-2i k_1}\right]
\end{equation}
and the diagonal $AA/BB$ hopping is given by
\begin{equation}
    \zeta_{\mathbf{k}} = 2t\cos(k_1) + 2t'\cos(k_2).
\end{equation}
Now for the {\120deg} phase, the $6\times6$ Hamiltonian reads
\begin{equation}
    H_{120}(\mathbf{k}) =
\begin{pmatrix}
\zeta_{\mathbf{k}}\mathbb{I}-\sigma\cdot \Delta_A & \delta^{AB}_{\mathbf{k}}\mathbb{I} & \delta^{AC}_{\mathbf{k}}\mathbb{I}\\
\delta^{*\,AB}_{\mathbf{k}}\mathbb{I} & \zeta_{\mathbf{k}}\mathbb{I}-\sigma\cdot \Delta_B & \delta^{BC}_{\mathbf{k}}\mathbb{I}\\
 \delta^{*\,AC}_{\mathbf{k}}\mathbb{I} & \delta^{*\,BC}_{\mathbf{k}}\mathbb{I} & \zeta_{\mathbf{k}}\mathbb{I}-\sigma\cdot \Delta_C
\end{pmatrix},
\end{equation}
with off-diagonal elements given by
\begin{equation}
    \delta^{AB}_{\mathbf{k}} = t\left[1 + e^{-i k_2} + e^{i(k_1-k_2)}\right],
    \quad
    \delta^{AC}_{\mathbf{k}} = t\left[1 + e^{-i k_2} + e^{-ik_1}\right],
    \quad
    \delta^{BC}_{\mathbf{k}} = t\left[1 + e^{-i k_1} + e^{i(k_2-k1)}\right].
\end{equation}
and diagonal elements given by
\begin{equation}
    \zeta_{\mathbf{k}} = 2t' \left[ \cos(k_1) + \cos(k_2) + \cos(k_1-k_2) \right].
\end{equation}

\end{document}